\PassOptionsToPackage{dvipsnames}{xcolor}
\documentclass[%
  aip,%
 amsmath,amssymb,floatfix,
reprint,%
twocolumn,%
jcp,%
longbibliography,
nofootinbib,
]{revtex4-1} 

\usepackage[normalem]{ulem}
\usepackage[all]{xy}
\usepackage{xcolor}   

\usepackage{mathtools}

\usepackage{amsmath,amsthm,latexsym,amssymb,amsfonts,epsfig}

\usepackage[english]{babel}
\usepackage[utf8]{inputenc}			
\usepackage{graphicx, texdraw}  
\usepackage{color}
\usepackage{bm}

\usepackage{mathtools}
\usepackage{mathrsfs} 

\usepackage[outer=1.3cm, inner=1.3cm, top=2cm, bottom=2cm]{geometry}  

\usepackage{enumerate}

\usepackage[colorlinks=true,citecolor=OliveGreen, linkcolor=blue]{hyperref}

\usepackage{csquotes} 

\usepackage{color}

\usepackage{amsmath}
\usepackage{amsfonts}
\usepackage{amssymb}

\newcommand{\vect}[1]{\bm{#1}}

\newcommand{\bv}{\vect{v}}

\newcommand{\G}{\mathcal{G}}
\newcommand{\R}{\vect{r}}
\newcommand{\Intd}{\mathrm{d }}

\newcommand{\vt}{\tilde{v}}

\newcommand{\Gt}{\tilde{\G}}

\newcommand{\bigO}{\mathcal{O}}

\newcommand{\bNabla}{\boldsymbol{\nabla}}

\newcommand{\vv}{{\boldsymbol{v}}}

\newcommand{\eg}{{\it e.g.}}

\newcommand{\bRho}{\boldsymbol{\rho}}
\newcommand{\bQ}{\vect{q}}

\usepackage[normalem]{ulem}

\newcommand{\fig}[1]{Fig.~\ref{#1}}

\newcommand{\tbl}[1]{Tab.~\ref{#1}}

\renewcommand{\vec}[1]{\boldsymbol{#1}}

\newcommand{\textmds}[1]{\textmd{\tiny{#1}}}

\usepackage{array}
\newcolumntype{L}[1]{>{\raggedright\let\newline\\\arraybackslash\hspace{0pt}}m{#1}}
\newcolumntype{C}[1]{>{\centering\let\newline\\\arraybackslash\hspace{0pt}}m{#1}}
\newcolumntype{R}[1]{>{\raggedleft\let\newline\\\arraybackslash\hspace{0pt}}m{#1}}

\newsavebox{\astrutbox}
\sbox{\astrutbox}{\rule[-5pt]{0pt}{20pt}}

\usepackage{setspace}

\usepackage{type1ec} %

\begin{document}

\title{State diagram of a three-sphere microswimmer in a channel}

\thanks{Article contributed to the Topical Issue of the Journal of Physics: Condensed Matter entitled \enquote{Transport in Narrow Channels} edited  by Paolo Malgaretti, Gleb Oshanin, and Julian Talbot.}

\author{Abdallah Daddi-Moussa-Ider}
\email{ider@thphy.uni-duesseldorf.de}

\affiliation
{Institut f\"{u}r Theoretische Physik II: Weiche Materie, Heinrich-Heine-Universit\"{a}t D\"{u}sseldorf, Universit\"{a}tsstra\ss e 1, 40225 D\"{u}sseldorf, Germany}

\author{Maciej Lisicki}

\affiliation
{Department of Applied Mathematics and Theoretical Physics, University of Cambridge, Wilberforce Rd, Cambridge CB3 0WA, United Kingdom}

\affiliation
{Institute of Theoretical Physics, Faculty of Physics, University of Warsaw, Pasteura 5, 02-093 Warsaw, Poland }

\author{Arnold J.T.M. Mathijssen}

\affiliation{Department  of  Bioengineering,  Stanford  University,
443  Via  Ortega,  Stanford,  CA  94305,  USA}

\author{Christian Hoell}

\affiliation
{Institut f\"{u}r Theoretische Physik II: Weiche Materie, Heinrich-Heine-Universit\"{a}t D\"{u}sseldorf, Universit\"{a}tsstra\ss e 1, 40225 D\"{u}sseldorf, Germany}

\author{Segun Goh}

\affiliation
{Institut f\"{u}r Theoretische Physik II: Weiche Materie, Heinrich-Heine-Universit\"{a}t D\"{u}sseldorf, Universit\"{a}tsstra\ss e 1, 40225 D\"{u}sseldorf, Germany}

\author{Jerzy B\l awzdziewicz}


\affiliation
{Department of Mechanical Engineering, Texas Tech University, Lubbock, TX 79409, USA}

\author{Andreas M. Menzel}

\affiliation
{Institut f\"{u}r Theoretische Physik II: Weiche Materie, Heinrich-Heine-Universit\"{a}t D\"{u}sseldorf, Universit\"{a}tsstra\ss e 1, 40225 D\"{u}sseldorf, Germany}

\author{Hartmut Löwen}
\email{hlowen@hhu.de}

\affiliation
{Institut f\"{u}r Theoretische Physik II: Weiche Materie, Heinrich-Heine-Universit\"{a}t D\"{u}sseldorf, Universit\"{a}tsstra\ss e 1, 40225 D\"{u}sseldorf, Germany}

\begin{abstract}

Geometric confinements are frequently encountered in soft matter systems and in particular significantly alter the dynamics of swimming microorganisms in viscous media.
Surface-related effects on the motility of microswimmers can lead to important consequences in a large number of biological systems, such as biofilm formation, bacterial adhesion and microbial activity.
On the basis of low-Reynolds-number hydrodynamics, we explore the state diagram of a three-sphere microswimmer under channel confinement in a slit geometry and fully characterize the swimming behavior and trajectories for neutral swimmers, puller- and pusher-type swimmers.
While pushers always end up trapped at the channel walls, {neutral swimmers} and pullers may further perform a gliding motion and maintain a stable navigation along the channel.
We find that the resulting dynamical system exhibits a supercritical pitchfork bifurcation in which swimming in the mid-plane becomes unstable beyond a transition channel height while two new stable limit cycles or fixed points that are symmetrically disposed with respect to the channel mid-height emerge.
Additionally, we show that an accurate description of the averaged swimming velocity and rotation rate in a channel can be captured analytically using the method of hydrodynamic images, provided that the swimmer size is much smaller than the channel height.

\end{abstract}
\date{\today}

\maketitle

\setstretch{1.}

\section{Introduction}\label{sec:introduction}

Microorganisms, particularly bacteria, constitute the bulk of the biomass on Earth and outnumber any other creatures. Despite their vast biological diversity and specific interaction with their environment, the physics of microscale fluid dynamics provides a unifying framework of the understanding of some aspects of their behavior\cite{purcell77,Lauga2009, elgeti15, marchetti13}. Swimming on the microscale is conceptually very different from the everyday macroscale experience\cite{lauga2016ARFM,zottl16, bechinger16}. Since the typical sizes and velocities of microswimmers are of the order of microns and microns per second, the Reynolds number characterizing the flow is {$\operatorname{Re} \ll 1$.} In this case, inertial effects can be disregarded compared to viscous effects in flow, and the motion of the fluid is described by linear Stokes hydrodynamics\cite{happel12, kim13}. This has a pronounced effect on the physiology and swimming strategies of microswimmers \cite{brennen77,lauga16} which have to comply to the limitations imposed by the time reversibility of Stokes flows, termed the scallop theorem by Purcell\cite{purcell77}.

One of the ways to overcome this barrier is to perform non-reciprocal swimming strokes. This can be achieved in systems of artificial biomimetic swimmers by introducing only few degrees of freedom, sufficient to gain propulsion but simplistic enough to remain analytically tractable. A well-known model example is the three-sphere swimmer designed by Najafi and Golestanian\cite{najafi04}. It encompasses three aligned spheres the mutual distance of which can be varied periodically in a controlled way. This guarantees the breaking of kinematic reversibility and leads to net translation along the axis of the body \cite{Najafi2005,Golestanian2008a,Golestanian2008b,golestanian08epje,alexander09}. The strength of this design lies in the possible experimental realizations involving colloids trapped in optical tweezers\cite{Leoni2009, grosjean16}. Similar bead-model designs have been proposed involving elastic deformations of one or both of the arms \cite{felderhof14, Montino2015,Montino2017,pande15,babel16,pande17, hosaka17, pande17njp}, non-collinear conformations leading to rotational motion \cite{Ledesma2012,Dreyfus2005,tenHagen15,kuchler16,Rizvi2018}, or new models with complex swimmer bodies and external propulsion forces \cite{degraaf16a, mathijssen2018universal}. 
A simple model for free-swimming animalcules composed of beads, subject to periodic forces has further been considered \cite{Felderhof2006, vladimirov13}.
Fascinating spatiotemporal patterns and unusual macroscopic rheological signatures arise from the interaction of numerous microswimmers, including the onset of collective and cohesive motion~\cite{gregoire04, mishra10, menzel12, heidenreich11, saintillan18}, emergence of dynamic clusters~\cite{goff16, scholz18}, laning~\cite{menzel13, kogler15, romanczuk16, menzel16njp} and wave patterns~\cite{ liebchen16, liebchen17, hoell17, liebchen17b}, motility-induced phase separation~\cite{tailleur08, buttinoni13, palacci13, speck14, speck15} and active turbulence~\cite{wensink12pnas, wensink12, dunkel13, heidenreich14, kaiser14, heidenreich16, lowen16}.

One of the main challenges of microfluidics has been to design and control the motion of fluids in microchannels, where the effects of confinement dominate the dynamics\cite{Squires2005, whitesides06}. The long-ranged nature of hydrodynamic interactions in low Reynolds number flows under geometrical confinement significantly influences the dynamics of suspended particles or organisms \cite{happel54}. Close confinement, e.g. in channels, can lead to a drastic increase in the range of interactions\cite{cui02, misiunas15}. Thus surface effects have to be accounted for when designing microfluidic systems \cite{stone04,lauga07} and affect translational and rotational mobilities of colloidal particles diffusing near boundaries \cite{lisicki14, lisicki16, rallabandi17b, daddi17, daddi17b, daddi17c, rallabandi17, daddi17d}. 
In living systems, walls have been demonstrated to drastically change the trajectories of swimming bacteria, such as {\it E. coli}~\cite{frymier95, diluzio05, lauga06,berke08, drescher11, mino11,ishimoto14, mathijssen16prl, lushi17, dauparas18},  or algae~\cite{ishimoto13, contino15}. As seen already in simplistic models involving two linked spheres near a wall\cite{dunstan2012}, a surprisingly rich behavior emerges, with the presence of trapping states, escape from the wall and non-trivial steady trajectories above the surface. This behavior has also been seen in an analogous system of self-phoretic active Janus particles\cite{spagnolie12, Uspal2015a,Uspal2015b, simmchen16, ibrahim15, mozaffari16, popescu16, ruhle17, mozaffari18}, where a complex phase diagram has been found, based on the initial orientation and the distance separating the swimmer from the wall. 
Additional investigations have considered the hydrodynamic interactions between two squirmers near a boundary\cite{li14}, the dynamics of active particles near a fluid interface~\cite{pimponi16, dominguez16, dietrich17}, swimming in a confining microchannel~\cite{elgeti09, zhu13, zottl12, zottl13, najafi13, bilbao13,yang16, elgeti16, liu16prl, wioland16,degraaf16understanding, wu16}, inside a spherical cavity~\cite{de-graaf17, reigh17,reigh17prf},  near a curved obstacle~\cite{spagnolie15, desai18} and in a liquid film \cite{mathijssen16jfm, dauparas16, mathijssen18}.
Meanwhile, other studies have considered the low-Reynolds number locomotion in non-Newtonian fluids~\cite{lauga07viscoelastic, pak12, zhu12, curtis13, riley14, datt15, mathijssen16,gomez16, natale17} where boundaries have been found to drastically alter the swimming trajectories of microswimmers.~\cite{yazdi14,yazdi15, yazdi17}.

The analysis of dynamics of a single model swimmer interacting with a boundary is a crucial first step towards the understanding of complex collective processes involving living systems close to boundaries. In this paper, we address theoretically and numerically the low-Reynolds-number locomotion of a linear three-sphere microswimmer in a channel between two parallel walls. We show that the swimmer flow signature (pusher, puller, {neutral swimmer}) determines its general behavior and explore the resulting phase diagrams discerning between the gliding, sliding and trapping modes of motion.

\begin{figure}
	\begin{center}
		\includegraphics[scale=0.62]{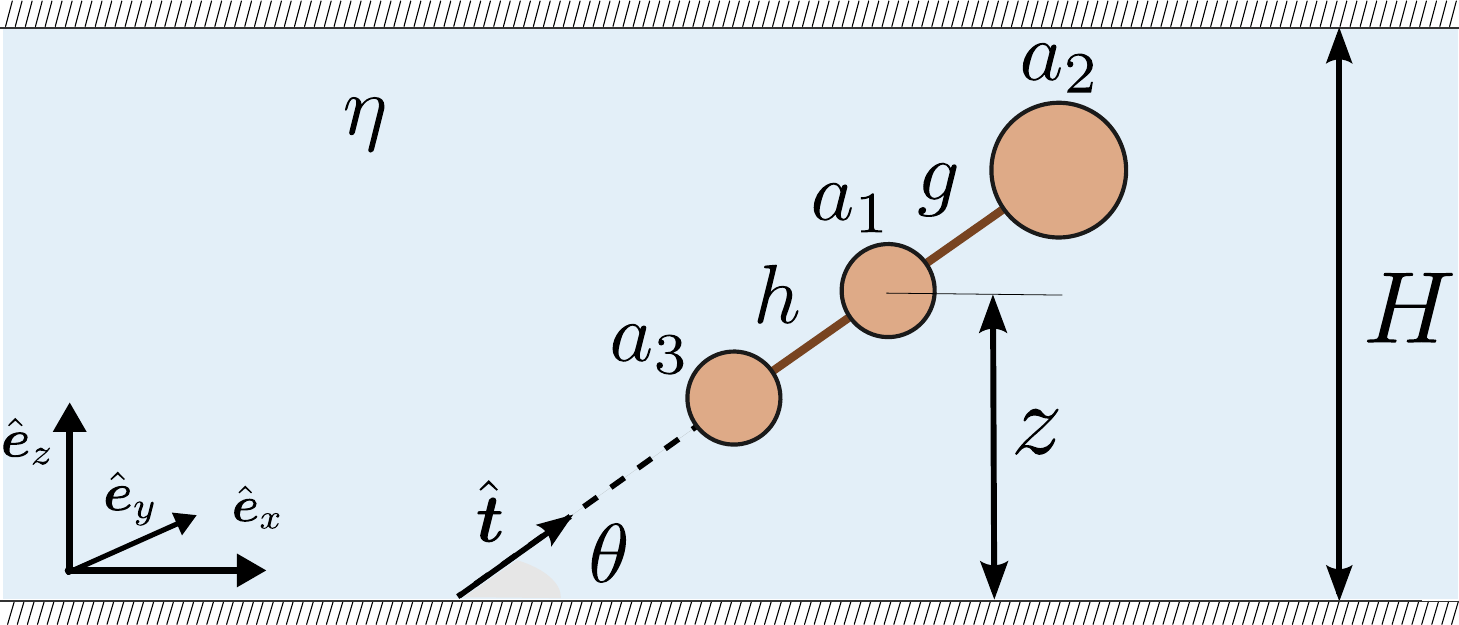}
		\caption{Illustration of a linear three-sphere microswimmer moving in a dimensional channel of constant height $H$. 
		The swimmer is directed along the unit vector~$\vect{\hat{t}}$ making an angle $\theta$ relative to the horizontal direction. 
		The central, front, and aft spheres composing the swimmer have different radii $a_1$, $a_2$, and $a_3$, respectively.
		The instantaneous positions of the front and aft spheres relative to the central sphere are denoted by $g$ and $h$, respectively.
		The vertical position of the swimmer is defined by the height of the central sphere~$z$ above the bottom wall.
		The fluid filling the channel is quiescent and characterized by a dynamic viscosity~$\eta$.}
		\label{Illustration}
	\end{center}
\end{figure}

The remainder of the paper is organized as follows. 
In Sec.~\ref{sec:theoreticalModel}, we introduce the model microswimmer and derive the swimming kinematics in a channel between two planar walls in the framework of low-Reynolds-number hydrodynamics.
We then present in Sec.~\ref{sec:stateDiagram} a state diagram representing the various swimming scenarios for a neutral three-sphere swimmer and present a simplified analytical model valid in the limit where the swimmer length is small compared to the channel height.
We discuss in Sec.~\ref{sec:pullerPusher} the behavior of puller- and pusher-type swimmers, finding that the former can maintain a stable navigation along the channel, while the latter inevitably ends up trapped at the channel walls.
We then examine the swimming stability about the mid-plane and show that a supercritical pitchfork bifurcation occurs beyond a certain transition channel height at which swimming at the centerline becomes unstable.
Concluding remarks and summary are provided in Sec.~\ref{sec:conclusions} and technical details are contained in Appendices~\ref{appendix:greensFcts} through \ref{appendix:formulas}.

\section{Theoretical model}\label{sec:theoreticalModel}

\subsection{Hydrodynamics background}

In low-Reynolds-number hydrodynamics, the flow is viscosity-dominated and the fluid motion is governed by the steady Stokes equations~\cite{happel12} 
\begin{subequations}   \label{stokesEq}
	\begin{align}
	\eta \bNabla^2 \bv(\R) - \bNabla P(\R) + \vect{f}_\mathrm{B} (\R)  &= 0 \, , \label{stokesEq_1} \\
	\bNabla \cdot \bv (\R) &= 0 \, ,  \label{stokesEq_2}
	\end{align} %
\end{subequations}
where $\eta$ denotes the fluid dynamic viscosity, and $\vect{v}(\R)$ and $P(\R)$ are respectively the fluid velocity and pressure fields at position $\R = (x,y,z)$ due to a bulk force density $\vect{f}_\mathrm{B}(\R)$ acting on the fluid by the immersed objects.

For a point-force singularity $\vect{f}_\mathrm{B} (\R) = \vect{f}  \delta(\R-\R_0)$ acting at position~$\R_0$ {in an otherwise quiescent fluid,} the solution for the induced velocity field and pressure is expressed in terms of the Green's functions
\begin{equation}
	v_i (\R) = \mathcal{F}_{ij} (\R,\R_0) f_j \, ,  \quad
	P (\R) = \mathcal{P}_{j} (\R,\R_0) f_j \, , 
	\label{velocityInTermsOfGreensFcts}
\end{equation}
where repeated indices are summed over following Einstein's convention. In the absence of confining boundaries, the fundamental solution is the Oseen tensor, $\boldsymbol{\mathcal{F}}=(8\pi\eta s)^{-1}(\boldsymbol{I} + \boldsymbol{s}\boldsymbol{s}/s^2)$ with $\boldsymbol{s}=\R-\R_0$ and $s=|\vect{s}|$. The solution for an arbitrary force distribution can  then be constructed by linear superposition.

The Green's functions in a channel between two parallel planar walls was first derived by Liron and Mochon~\cite{liron76} using the image technique and a Fourier transform. 
In Appendix~\ref{appendix:greensFcts}, we present a modified approach based on decomposing the Fourier-transformed vector fields into their longitudinal, transverse, and normal components.
Upon inverse Fourier transformation, the Green's functions can then be expressed in terms of Bessel integrals of the first kind.  
Alternatively, following the method by Mathijssen \textit{et al.}~\cite{mathijssen16jfm}, the Green's function may be expressed as an infinite series of image reflections.
In Appendix~\ref{appendix:Arnold} we derive the recursion relations that yield the successive image systems, and provide explicit expressions for these. 
Truncation of this series can prove to be computationally advantageous, provided a suitable number of images is chosen.
In the limiting case of an infinitely wide channel, both Green's functions reduce to the familiar Oseen tensor.
{The image reflection method has previously been employed to address the behavior of swimming bacteria near a hard surface~\cite{berke08} or an air-fluid interface~\cite{di-leonardo11}.}

\subsection{Swimmer dynamics}

In the following, we consider the motion of a neutrally buoyant swimmer in a fluid bounded by two parallel planar walls infinitely extended in the planes $z=0$ and $z=H$. 
As a model swimmer, we employ the linear three-sphere microswimmer originally proposed by Najafi and Golestanian~\cite{najafi04}.
The simplicity of the model provides a handy framework that allows a direct investigation of many aspects in low-Reynolds-number locomotion.
The swimmer is composed of three spheres {of radii $a_1$ (central), $a_2$ (front), and $a_3$ (rear)} arranged colinearly via dragless rods.
The periodic changes in the mutual distances between the spheres are set to perform a non-reversible sequence leading to propulsive motion (see Fig.~\ref{Illustration} for an illustration of the model swimmer moving in a channel between two walls.)

The instantaneous orientation of the swimmer relative to the channel walls is described by the two-dimensional unit vector $\vect{\hat{t}} = \cos\theta \, \vect{\hat{e}}_x + \sin\theta \, \vect{\hat{e}}_z$ directed along the swimming axis.
Under the action of the internal forces acting between the spheres, actuated, e.g., by embedded motors, the lengths of the rods connecting the spheres change periodically around mean values.
Specifically,
\begin{equation}
	\R_1-\R_3 = h(t) \vect{\hat{t}} \, , \quad 
	\R_2-\R_1 = g(t) \vect{\hat{t}} \, , 
	\label{prescribedMotion}
\end{equation}
where $h(t)$ and $g(t)$ are periodic functions prescribing the instantaneous mutual distances between adjacent spheres, which we choose to be harmonic,
\begin{subequations}
	\begin{align}
		g(t) &= L_1 + u_{10} \cos \left(\omega t\right) \, , \\
		h(t) &= L_2 + u_{20} \cos \left(\omega t + \delta\right) \, , 
	\end{align}
\end{subequations}
where $\omega$ is the oscillation frequency of motion and $\delta \in [0,2\pi)$ is a phase shift necessary for the symmetry breaking.
Here, $L_1$ and $L_2$ stand for the mean arm length connecting the central sphere to the front and aft spheres, respectively.
In addition, $u_{10}$ and $u_{20}$ are the corresponding amplitudes of oscillation.
Unless otherwise stated, we will consider consistently throughout this manuscript that $L_1=L_2 =:L$ and $u_{10}=u_{20} =:u_0$.
{We further mention that the sphere radii and the oscillation amplitudes should be chosen small enough in such a way that the inequalities $a_1+a_2+2|u_{0}| \ll L$ and $a_1+a_3+2|u_{0}| \ll L$ remain satisfied.}
Moreover, we scale from now on all the lengths by~$L$ and the times by~$\omega^{-1}$.

We now briefly outline the main steps involved in the derivation of the swimming velocity and inclination.
In Stokes hydrodynamics, the suspended particles take instantaneously on the velocity of the embedding flow since inertial effects are negligible.
Additionally, the translational velocities of the three spheres are linearly related to the internal forces acting on them via, 
\begin{equation}
	\vect{V}_\gamma = \frac{\Intd \R}{\Intd t} = \sum_{\lambda=1}^{3} \bm{\mu}^{\gamma\lambda} \cdot \vect{f}_\lambda \, ,
	\label{relateVandFviaMobility}
\end{equation} 
where $\bm{\mu}^{\gamma\lambda}$ denotes the hydrodynamic mobility tensor bridging between the translational velocity of sphere  $\gamma$ and the force exerted on sphere $\lambda$.
The mobility tensor is symmetric positive definite~\cite{swan07} and encompasses the effect of many-body hydrodynamics interactions. In this work, however, for the sake of simplicity we consider only contributions  stemming from the hydrodynamic interaction between pairs of particles $(\gamma \ne \lambda)$, in addition to contributions relative to the same particle $(\gamma = \lambda)$ designated as self mobility functions~\cite{kim13}.

Taking the time derivative with respect to the laboratory frame on both sides of Eq.~\eqref{prescribedMotion} yields
\begin{subequations}\label{V2UndV3vontimeDerivative}
	\begin{align}
		\vect{V}_2 &= \vect{V}_1 + \dot{g} \, \vect{\hat{t}}    + g \dot{\theta} \, \vect{\hat{n}} \, , \label{V2vontimeDerivative} \\ 
		\vect{V}_3 &= \vect{V}_1 - \dot{h} \, \vect{\hat{t}}    - h \dot{\theta} \, \vect{\hat{n}} \, , \label{V3vontimeDerivative}
	\end{align} 
\end{subequations}
wherein dot stands for a derivative with respect to time.
Moreover, $\vect{\hat{n}} = -\sin\theta \, \vect{\hat{e}}_x + \cos\theta \, \vect{\hat{e}}_z$ is a unit vector perpendicular (rotated 90 degrees anticlockwise) to the unit vector~$\vect{\hat{t}}$.
Accordingly, the triplet $(\vect{\hat{e}}_y, \vect{\hat{n}}, \vect{\hat{t}})$ forms a direct orthonormal basis in the frame of reference associated with the swimmer.

For the determination of the unknown internal forces acting between the spheres, a total of six equations is required.
By projecting Eqs.~\eqref{V2UndV3vontimeDerivative} onto the orientation vector $\vect{\hat{t}}$, two scalar equations are readily obtained.
Projecting these equations onto the normal direction $\vect{\hat{n}}$ and eliminating the rotation rate yields an additional equation.
Three further scalar equations are obtained by enforcing the physical constraint that the swimmer does not exert a net force or torque on the surrounding fluid. Specifically
\begin{equation}
	\sum_{\lambda=1}^{3} \vect{f}_\lambda = 0 \, , \qquad
	\sum_{\lambda=1}^{3} \left(\R_\lambda-\vect{r}_0 \right) \times \vect{f}_\lambda = 0\, , 
\end{equation}
where $\times$ stands for the cross (outer) product and $\R_0$ denotes an arbitrary reference point, which we choose to be the position of the central sphere $\R_1$. 
The internal forces acting between the spheres follow from solving the resulting system of six linearly independent equations using the standard substitution technique.

In order to investigate the swimming behavior, we choose to follow the trajectory of the central sphere whose velocity can readily be determined from \eqref{relateVandFviaMobility} upon knowledge of the internal forces.
The instantaneous rotation rate of the swimmer can then be calculated from
\begin{equation}
	\dot{\theta} = \frac{1}{g} \left( \vect{V}_2 - \vect{V}_1 \right) \cdot \vect{\hat{n}}
				 = \frac{1}{h} \left( \vect{V}_1 - \vect{V}_3 \right) \cdot \vect{\hat{n}} \, .
\end{equation}


\section{Swimming state diagram}\label{sec:stateDiagram}

\subsection{Behavior near a single wall}

Having outlined the general procedure for the determination of the equations governing the swimmer dynamics, we next derive approximate expressions for the swimming translational and rotational velocities.
We firstly consider the limiting case of an infinitely wide channel $H \to\infty$ and derive the averaged equations of motion for a swimmer located at a finite distance above a single wall infinitely extended in the plane $z=0$.
In addition, we restrict our attention to the particular case where the spheres have the same radius~$a$ as originally proposed in the Najafi and Golestanian design~\cite{najafi04}.
The general case for arbitrary particle radius will be discussed in the following section.

The Green's functions satisfying the no-slip boundary condition at an infinitely extended hard wall are expressed in the form of the Blake tensor~\cite{blake71} providing the leading-order terms in the pair hydrodynamic interactions.
Restricting ourselves for simplicity to the point-particle framework, the scaled self-mobility functions for a sphere located at height~$z$ above a rigid wall are given up to $\bigO \left( \left(a/z\right)^3 \right)$ by~\cite{happel12}
\begin{equation}
	\frac{\mu_\parallel}{\mu_0} =  1-\frac{9}{16} \frac{a}{z}   \, , \qquad
	\frac{\mu_\perp}{\mu_0} =  1-\frac{9}{8} \frac{a}{z}   \, , 
	\label{selfMobilitySingleWallLeadingOrder}
\end{equation}
for the translational motion parallel and perpendicular to the wall, respectively.
Here $\mu_0=\left(6\pi\eta a\right)^{-1}$ denotes the usual bulk mobility given by the Stokes law. {(In our simulations, however, we use more detailed predictions obtained by the method of reflections incorporating nine images, and described in detail in Appendix \ref{appendix:Arnold}.)}

By performing a Taylor series expansion up to $\bigO (a^3)$ of the swimming velocity and rotation rate, the approximate differential equations governing the swimming dynamics above a single wall, averaged over one oscillation period, can be presented in the form
\begin{subequations}\label{swimmingTrajABC_neutral}
	\begin{align}
		\frac{\Intd x}{\Intd t} &= V_0 + K A(z) \, , \label{swimmingTrajA_neutral}\\
		\frac{\Intd z}{\Intd t} &= \big( V_0 + K B(z) \big) \theta \, , \label{swimmingTrajB_neutral} \\
		\frac{\Intd \theta}{\Intd t} &= K C(z) \, , \label{swimmingTrajC_neutral}
	\end{align} 
\end{subequations}
where we have assumed small inclination angles relative to the horizontal direction such that $\sin\theta \sim \theta$ and $\cos\theta \sim 1$.
Moreover,
\begin{equation}
	V_0 = -\frac{aK}{24} \left( 7+5a \right) \, , \label{bulkSwimmingVelocity}
\end{equation}
is the bulk swimming speed in the absence of a boundary, and
\begin{equation}
	K:= \langle g\dot{h} - h\dot{g} \rangle 
	= -u_{10} u_{20} \sin \delta 
	=  -u_0^2 \sin \delta  \, .
\end{equation}
Here $\langle \cdot \rangle$ stands for the time-averaging operator over one full swimming cycle, defined by
\begin{equation}
	\langle \cdot \rangle := \frac{1}{2\pi} \int_0^{2\pi} (\cdot) \, \Intd t \, .
\end{equation}
Evidently, a net motion over one swimming cycle occurs only if the phase shift $\delta \notin \{0,\pi\}$.
In the remainder of this article, we take $\delta=\pi/2$ for which the swimming speed is maximized.

In addition, $A, B,$ and $C$ are highly nonlinear functions of~$z$ which are explicitly provided to leading order in~$a$ in Appendix~\ref{appendix:formulas}.
In the far-field limit, in which the distance separating the swimmer from the wall is very large compared to the swimmer size $(z\gg 1)$, these functions up to~$\bigO\left( z^{-5} \right)$ read
\begin{subequations}\label{ABC_GleichungenFar}
	\begin{align}
		A(z) &= -\frac{287}{1024} \frac{a^2}{z^3} \, , \\
		B(z) &= \left( \frac{21}{64}-\frac{77a}{256} \right) \frac{a}{z^3} \, , \\
		C(z) &=  \, \frac{135}{1024} \frac{a(2+3a)}{z^4} \, .
	\end{align} 
\end{subequations}
Remarkably, the leading-order term in the wall-induced correction to the swimming velocity decays in the far field as~$z^{-3}$.
Not surprisingly, the dipolar contribution (decaying as~$z^{-2}$) induced by a three-sphere microswimmer vanishes if the front and aft spheres have equal radii (see Appendix~\ref{appendix:3sphereSwimmer}).
As a result, the leading-order in the velocity flow field possesses a quadrupolar flow structure that decays as inverse cube of distance.
Approximate swimming trajectories are readily obtained by integrating Eqs.~\eqref{swimmingTrajABC_neutral} for given initial orientation and distance from the wall.


\subsection{Approximate swimming trajectories in a channel}

We next shift our attention to the swimming motion in a channel bounded by two parallel infinitely extended walls.
As already pointed out, an accurate description of the channel-mediated hydrodynamic interactions requires utilization of the Green's functions that satisfy the no-slip boundary conditions at both walls simultaneously.
This approach, however, involves improper (infinite) integrals whose numerical evaluation at every time step is computationally expensive.
In order to overcome this difficulty, we use as an alternative framework the successive image reflection technique. 
The latter consists of generating an infinite series of images containing Stokeslets and higher-order derivatives that satisfy the no-slip boundary conditions on both walls asymptotically.
Further technical details on the derivation of the flow field using multiple reflections are provided in Appendix~\ref{appendix:Arnold}.
Throughout this work, a total of eight reflections is consistently employed for the numerical evaluation of the Green's functions.


In order to proceed analytically, we restrict ourselves for simplicity to the first two image systems following Oseen's classical approximation~\cite{oseen28}.
This approach suggests that the wall-induced corrections to the hydrodynamic interactions between two planar parallel rigid walls could conveniently be approximated by superposition of the contributions stemming from each single wall independently.
Accordingly, it follows from Eqs.~\eqref{swimmingTrajABC_neutral} that the averaged swimming velocities in a channel between two walls can adequately be approximated as
\begin{subequations}\label{swimmingVelocityTwoReflections}
	\begin{align}
		\frac{\Intd x}{\Intd t} &= V_0 + K \left( A(z) + A(H-z) \right) \, , \\
		\frac{\Intd z}{\Intd t} &= \big( V_0 + K \left( B(z) + B(H-z) \right) \big) \theta  \, , \label{diffZdiffT} \\
		\frac{\Intd \theta}{\Intd t} &= K \big( C(z) - C(H-z) \big) \, , \label{diffThediffT}
	\end{align} 
\end{subequations}
where again the inclination angle is assumed to vary within a narrow range relative to the horizontal direction.

\begin{figure}
\begin{center}
\includegraphics[scale=0.75]{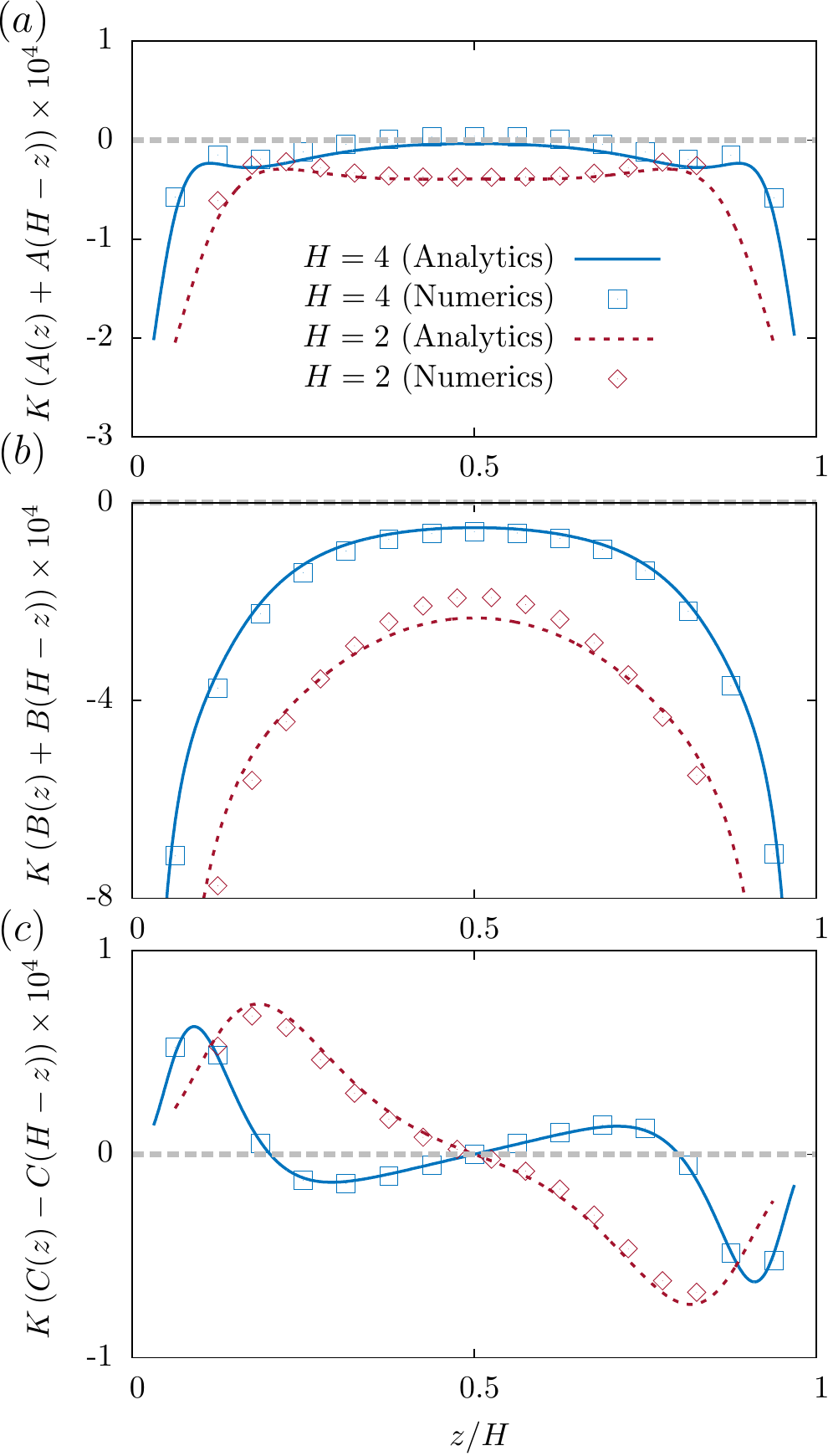}
\end{center}
\caption{(Color online) $(a)$~and~$(b)$ {Channel-induced corrections to the translational swimming velocities} along the $x$ and $z$ directions, respectively, and $(c)$ rotation rate versus the vertical distance~$z$ about $\theta\sim 0$.
The analytical expressions based on the superposition approximation given by Eqs.~\eqref{swimmingVelocityTwoReflections} derived up to $\bigO(a^3)$ are shown as dashed and solid lines for $H=2$ and $H=4$, respectively.
Symbols are the numerically exact results obtained using a total of eight reflections for $H=2$ (diamonds) and $H=4$ (squares).
Horizontal (gray) dashed lines are the corresponding bulk values.
Here we consider a neutral swimmer with equal sphere radii $a=0.1$ and an amplitude of arm oscillations $u_0=0.1$. 
}
\label{ABC_Fig}
\end{figure}

In Fig.~\ref{ABC_Fig}, {we show the channel-induced corrections} to the swimming velocities and rotation rate as functions of the vertical distance~$z$ for a neutral swimmer of equal sphere radii $a =0.1$.
The simplistic superposition approximation given by Eqs.~\eqref{swimmingVelocityTwoReflections} is shown as dashed and solid lines for channel heights $H=2$ and $H=4$, respectively.
The corresponding numerical solutions obtained using a total of eight reflections are shown as symbols, where diamonds and squares correspond to $H=2$ and $H=4$, respectively.
Here we consider a small amplitude of oscillations $u_0=0.1$.

{We observe that the corrections to the swimming velocities (Fig.~\ref{ABC_Fig}$(a)$-$(b)$) remain typically constant around the channel mid-height and mostly monotonically increase in magnitude in the proximity of the walls due to the increased drag exerted on the swimmer.}
{Upon decreasing the channel height, the drag force resulting from the resistance of the channel walls and opposing the motion through the fluid becomes more pronounced.
For instance, swimming in the mid-plane of a channel of height $H=2$ leads to increased drag of about $13~\%$ relative to the bulk value, while this increase is found to be of only about $2~\%$ for $H=4$.}
{The increase in the drag force for motion of arbitrary direction is mostly larger in the $z$ direction than in the $x$ direction since it is easier to move the fluid aside than to push it into or to squeeze it out of the gap between the swimmer and the channel walls.}

Since the vertical velocity scales linearly with the inclination angle (c.f.\@ Eq.~\eqref{diffZdiffT}), a swimmer that is initially aligned parallel to the walls and released from a height of vanishing rotation rate will undergo a purely gliding motion along the channel.
By examining the variations of the rotation rate {(Fig.~\ref{ABC_Fig}~$(c)$)} we observe that the evolution equations for the swimming trajectories display either one or three fixed points in the comoving frame translating parallel to the channel walls.
The first fixed point is trivial and occurs at the channel mid-height $(z/H=1/2)$ where both walls have the same effect on the orientation of the swimmer.
For $H=4$, two nontrivial fixed points symmetrically placed with respect to the channel mid-height are reached at $z/H \simeq 0.2$ and $z/H \simeq 0.8$.

The superposition approximation is found to be in a good agreement with the full numerical solution along the channel.
A small mismatch, notably for $H=2$ in the normal velocity (Fig.~\ref{ABC_Fig}~$(b)$), is a drawback of the approximations proposed here.
A good estimate of the swimming trajectories in a channel can therefore be made using the first two reflections provided that the swimmer size is much smaller that the channel height.

\begin{figure}
\begin{center}
\includegraphics[scale=0.85]{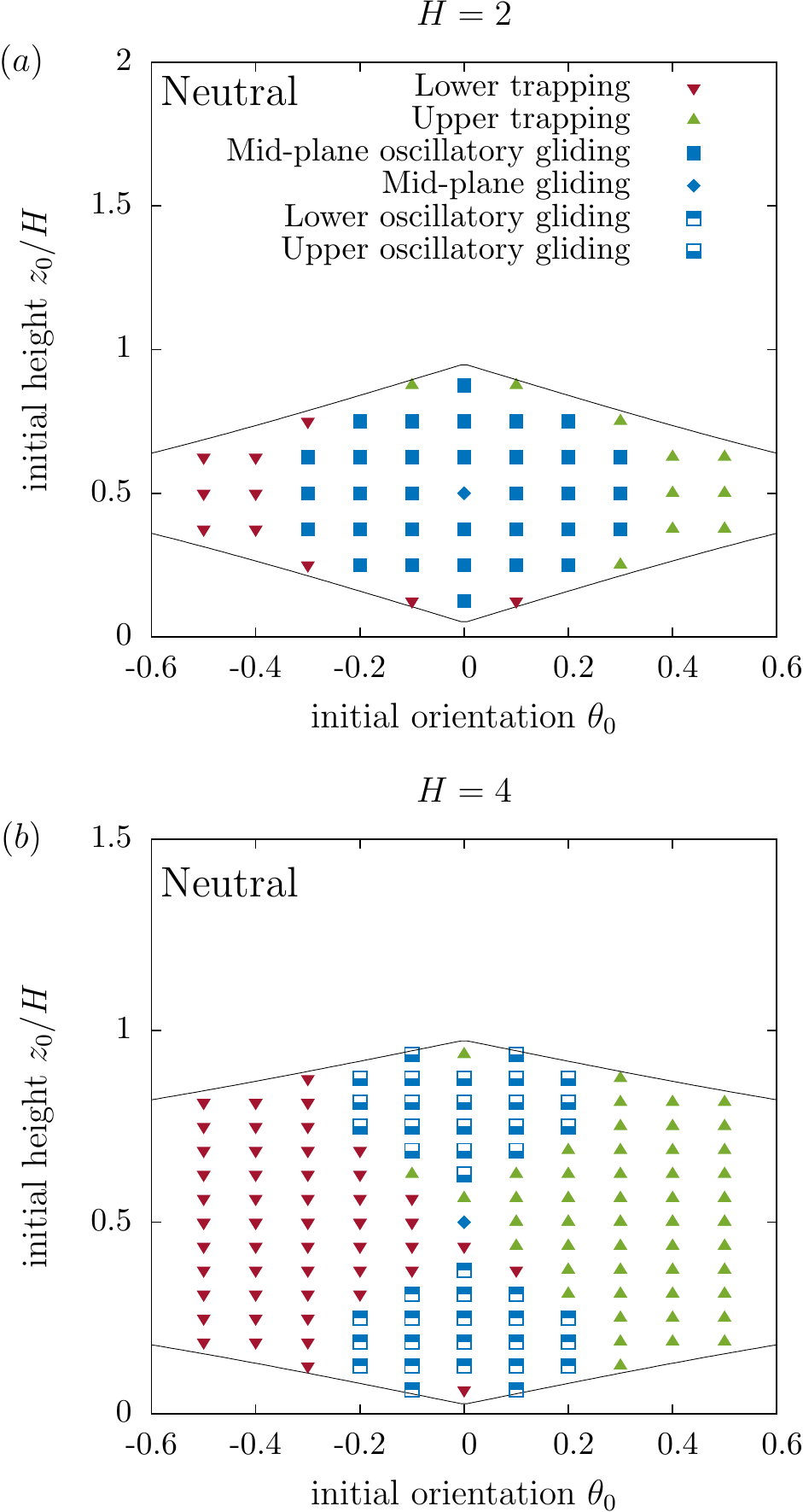}
\end{center}
\caption{(Color online) State diagram illustrating the swimming scenarios displayed by a neutral three-sphere swimmer of equal sphere radii $a=0.1$ confined in a channel between two parallel planar walls for $(a)$ $H=2$ and $(b)$ $H=4$.
Symbols represent the final swimming states for a given initial orientation and distance along the channel.
Downward pointing triangles (red) indicate trapping near the lower wall whereas upward pointing triangles (green) stand for trapping near the upper wall.
Boxes (blue) represent the oscillatory gliding state at the channel centerline while half-filled (blue) boxes correspond to the oscillatory gliding states above the corresponding wall.
A (blue) diamond marks the trivial perpetual motion along the exact centerline of the channel.
Solid lines correspond to forbidden situations in which one of the spheres is initially in contact with the channel walls.
Here we take an amplitude of oscillations $u_{0} = 0.1$. 
}
\label{Phase-diagram}
\end{figure}


\begin{figure}
\begin{center}
\includegraphics[scale=1]{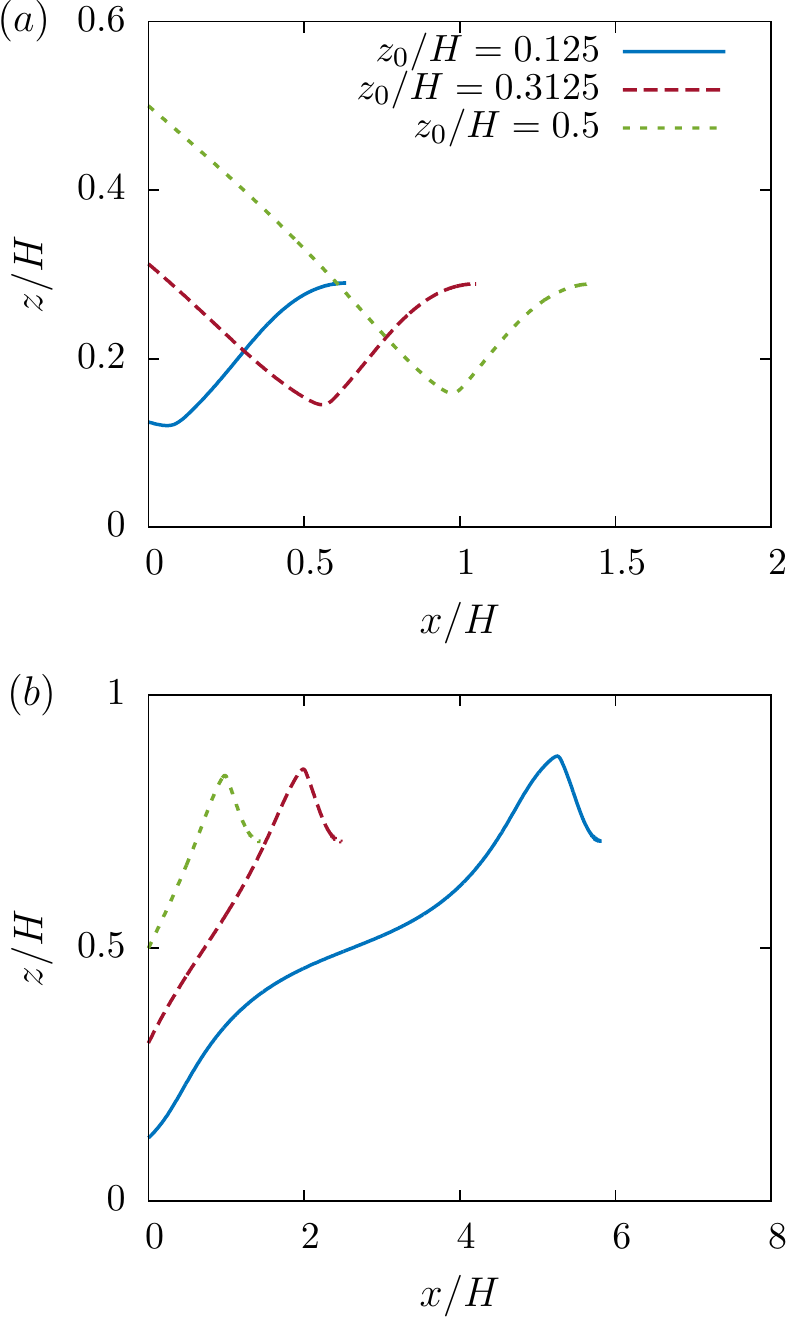}
\end{center}
\caption{
{(Color online) Typical swimming trajectories in $(a)$ the lower trapping for $\theta_0=-0.3$ and $(b)$ the upper trapping states for $\theta_0=0.3$ for various initial distances~$z_0$ in a channel of height $H=4$.
In the steady state, the swimmer ends up trapped by a wall and attains a stable hovering state at a constant height close to that wall.
This is why the trajectories end at a certain point.
}
}
\label{TrappingNeutralInset}
\end{figure}

Fig.~\ref{Phase-diagram} shows the state diagram displayed by a neutral three-sphere microswimmer of equal sphere radii, swimming in a channel for two different wall distances $(a)~H=2$ and $(b)~H=4$.
The state diagram is obtained by integrating the full nonlinear equations governing the swimmer dynamics numerically using a fourth-order Runge-Kutta scheme with adaptive time stepping~\cite{press92}.
{The hydrodynamic mobility functions used in the simulations are obtained using the method of reflection with a total of nine images, providing a good accuracy even at small sphere--wall distances, as compared to far-field representation.
A systematic comparison between the expressions of the self mobilities as obtained from the method of reflections and the exact multipole method~\cite{bhattacharya02, bhattacharya05jfm, bhattacharya05} is provided in the Supporting Information}
\footnote{See Supporting Information at [URL will be inserted by the editor] for approximate expressions of the self mobilities as obtained from the method of reflections in addition to a direct comparison with other approaches. }.
Depending on the initial orientation and distance along the channel, the swimmer may be trapped by either walls (downward and upward pointing triangles) or undergoes a nontrivial oscillatory gliding motion at a constant mean height either at the channel centerline (squares in Fig.~\ref{Phase-diagram}~$(a)$) or at a moderate distance near the channel wall (half-filled blue boxes in Fig.~\ref{Phase-diagram}~$(b)$).

A swimmer that is initially aligned parallel to the walls {$(\theta_0=0)$} and released from the trivial fixed point at the channel mid-height $(z_0=H/2)$ (blue diamond) undergoes a purely gliding motion without oscillations.
In the trapped state, the swimmer moves along a curved trajectory before it attains a hovering state during which the inclination angle approaches $\theta = -\pi/2$ for the lower trapping and $\theta = \pi/2$ for the upper trapping.
Only trapping occurs if the swimmer is sufficiently oriented away from the horizontal direction at varying extent depending upon the channel height.
{Fig.~\ref{TrappingNeutralInset} shows exemplary trajectories displayed by a neutral swimmer released from various initial heights with orientations $\theta_0=-0.3$ (for the lower trapping state) and $\theta_0=0.3$ (for the upper trapping states).
After a transient evolution, the swimmer reorients itself perpendicular to the nearest wall and reaches a stable hovering state at a separation distance of about $z \simeq 1.12$.
The final height is found to be independent of the initial inclination or distance from the wall in a way similar to that previously observed near a single boundary~\cite{daddi18}.
Physically, the hovering state corresponds to the situation in which the propulsion forces are equilibrated by the resistive viscous forces pushing the swimmer away from the nearest boundary. }


\begin{figure}
\begin{center}
\includegraphics[scale=1]{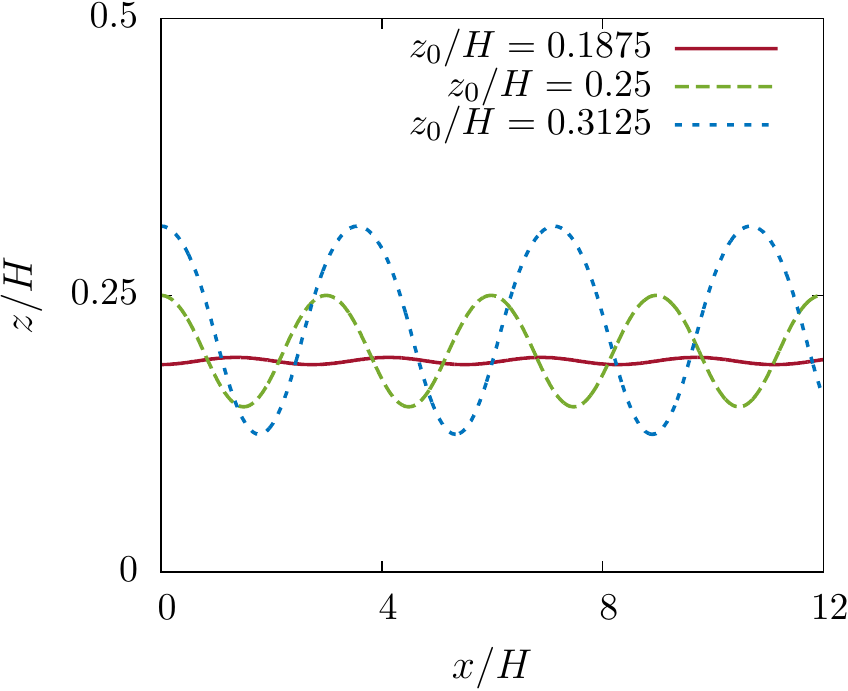}
\end{center}
\caption{(Color online) Exemplary swimming trajectories in {the lower oscillatory gliding state} for a separation $H=4$ between the walls.
The swimmer is initially aligned parallel to the walls $(\theta_0=0)$ and released from various initial distances~$z_0$. 
The amplitude of oscillations and frequency are strongly sensitive to the initial conditions.
A nearly  vanishing amplitude is observed {for $z_0/H \simeq 0.1875$} close to the stable fixed point in Fig.~\ref{ABC_Fig}~$(c)$.
The inclination angles show an analogous oscillatory behavior around a zero mean value.
Here we set $a=0.1$ {and an amplitude of arm oscillation $u_0=0.1$.}
}
\label{Oscillations}
\end{figure}

We show in Fig.~\ref{Oscillations} typical swimming trajectories in {the lower oscillatory gliding state} for a swimmer that is released from different initial heights along the parallel direction $(\theta_0=0)$ in a channel of height $H=4$.
In particular, the amplitude of oscillations almost vanishes {when $z_0/H \simeq 0.1875$} for which the swimmer undergoes a purely gliding motion at a constant height.
Not surprisingly, we have previously shown in Fig.~\ref{ABC_Fig}~$(c)$ that there exist in the comoving frame two nontrivial fixed points symmetrically placed relative to the channel centerline at $z/H \simeq 0.2$ and $z/H \simeq 0.8$ in addition to the trivial fixed point at the middle of the channel.
As the initial swimming location is shifted far away from the fixed points, the amplitude of oscillations grows gradually before the swimmer ends up trapped by the nearest wall.
{The swimmer shows an analogous behavior in the upper oscillatory state upon making the transformation $z \to H-z$ due to the system reflectional symmetry with respect to the channel mid-plane}
\footnote{{See Supporting Information at [URL will be inserted by the editor] for illustrative movies showing the swimming behaviors of a neutral three-sphere swimmer in a channel.
Movie~1 illustrates the lower trapping state $(z_0/H=0.125, \theta_0=-0.3)$ shown in Fig.~4$(a)$ (solid blue line).
Movie~2 illustrates the upper trapping state $(z_0/H=0.125, \theta_0=0.3)$ shown in Fig.~4$(b)$ (solid blue line).
Movie~3 shows the lower oscillatory gliding $(z_0=0.3125, \theta_0=0)$ presented in Fig.~5 (short-dashed blue line).
For illustrative purposes, the sizes of the spheres are not shown in real scale in the movies.}}.


\section{Swimming puller v.s. pusher}\label{sec:pullerPusher}

Having analyzed in detail the swimming behavior of a neutral three-sphere swimmer of equal sphere radii, we next consider the more general situation and allow for differently sized spheres for which the swimming stroke is not time-reversal covariant~\cite{alexander09}.
For that purpose, we introduce the radius ratios $r_{2} := a_2/a_1$ and $r_{3} := a_3/a_1$ and use $a$ to denote the radius of the central sphere $a_1$.
It should be noted that $r_2$ and $r_3$ must vary only in such a way that the {inequalities $(1+r_{2})a+2|u_0| \ll L$ and $(1+r_{3})a+2|u_0| \ll L$} remain satisfied during a full swimming cycle for the above-mentioned approximations to be valid.

In a bulk fluid, the flow field induced by a general three-sphere swimmer can conveniently be written in the far-field limit as a superposition of dipolar and quadrupolar flow fields (c.f.\@ Appendix~\ref{appendix:3sphereSwimmer}), whose coefficients are respectively given by 
\begin{equation}
	\alpha = \frac{3}{4} \frac{r_2 - r_3}{a} \, , \qquad
	\sigma = \frac{3}{56} \frac{4(r_2+r_3)-3}{a^2} \, , 
\end{equation}
where the swimmer is termed as pusher (extensile) if $\alpha>0$ as it then pushes out the fluid along its swimming axis, and as puller (contractile) if $\alpha<0$ as in that case it pulls in the fluid along its swimming path~\cite{Lauga2009}.
The swimmer studied in the previous section is a neutral {swimmer}, because $\alpha=0$, and the dominant contribution to the flow-far field thus is a quadrupole.

Keeping for convenience the same notation for the approximated swimming velocities and rotation rate as before, the averaged equations of motion of a general three-sphere swimmer near a single wall about the horizontal direction, can be presented up to $\bigO(a^3)$ as
\begin{subequations}
	\begin{align}
		\frac{\Intd x}{\Intd t} &= V_0 + K A(z) \, , \\
		\frac{\Intd z}{\Intd t} &= \big( V_0 + K B(z) \big) \theta + KD(z)   \, , \\
		\frac{\Intd \theta}{\Intd t} &= K C(z) \, , 
	\end{align}
\end{subequations}
where the bulk swimming velocity is now given by
\begin{eqnarray}
	V_0 = a \left( V_{10} + a V_{20} \right) \, . \label{bulkSwimmingVelocityGeneral}
\end{eqnarray}
The coefficients $V_{10}$ and $V_{20}$ are functions of $r_2$ and $r_3$ only.
They are explicitly given in appendix~\ref{appendix:formulas}.
In particular, $V_{10} = -\frac{7K}{24}$ and $V_{20} = -\frac{5K}{24}$ when $r_2=r_3$ directly leading to Eq.~\eqref{bulkSwimmingVelocity}.

In the far-field limit, the generalized expressions for the functions $A(z)$, $B(z),$ and $C(z)$ are 
\begin{subequations}\label{expr_ABC_general}
	\begin{align}
		A(z) &=   \frac{a^2 A_{23}}{z^3} \, , \label{expr_A_general} \\
		B(z) &= a \left(  \frac{B_{13}}{z^3}
				+  \frac{a B_{23}}{z^3} \right) \, , \label{expr_B_general} \\
		C(z) &=  a \left( C_{14} + a C_{24} \right) \frac{1}{z^4} \, . \label{expr_C_general}
	\end{align} 
\end{subequations}
In addition,
\begin{equation}
	D(z) = a (r_{3}-r_{2}) \left( \frac{D_{14}}{z^4} + \frac{a}{z^2} \left( D_{22} + \frac{D_{24}}{z^2}  \right) \right) \, . \label{expr_D_general} \\
\end{equation}
The coefficients $A_{ij}$, $B_{ij}$, $C_{ij},$ and $D_{ij}$ are provided in Appendix~\ref{appendix:formulas}.
{The far-field equations~\eqref{expr_ABC_general} reduce to Eqs.~\eqref{ABC_GleichungenFar} in the particular case of $r_{2}=r_{3}$.}

By accounting only for the leading order in $1/z$, the normal velocity in the flow-far field reads $\Intd z/\Intd t = a^2K(r_3-r_2)D_{22} z^{-2}$.
For a pusher-like swimmer $(r_2>r_3)$, it follows that $\Intd z/\Intd t < 0$, and thus the swimmer is expected to be trapped by the bottom wall by noting that $D_{22}<0$ and bearing in mind that $K<0$.
For a puller-like swimmer, however, $\Intd z/\Intd t > 0$ leading to an escape from the wall.
These observations are in agreement with previous studies indicating that a noiseless pusher swimming parallel to a wall will be attracted whereas a puller will be repelled~\cite{lopez14,schaar15}.
It is worth mentioning that the dipolar flow signature neither emerges in the $x$-component of the swimming velocity nor in the rotation rate.

By considering only for the first two image systems (superposition approximation), the generalized swimming velocities in a channel bounded by two walls can conveniently be approximated by
\begin{subequations}
	\begin{align}
		\frac{\Intd x}{\Intd t} &= V_0 + K \left( A(z) + A(H-z) \right) \, , \\
		\frac{\Intd z}{\Intd t} &=  \Big( V_0 +  K \left(B(z) + B(H-z)\right) \Big) \theta \notag \\
		&+ K \left( D(z)-D(H-z) \right)  \, , \label{diffZdiffTPushPull} \\
		\frac{\Intd \theta}{\Intd t} &= K \big( C(z) - C(H-z) \big) \, . \label{diffThediffTPushPull}
	\end{align}
\end{subequations}
Explicit analytical expressions for the functions $A$, $B$, $C$, and $D$ for a general three-sphere swimmer are rather complex and lengthy, and thus have not been listed here.

\subsection{State diagram in a channel}

\begin{figure}
\begin{center}
\includegraphics[scale=0.9]{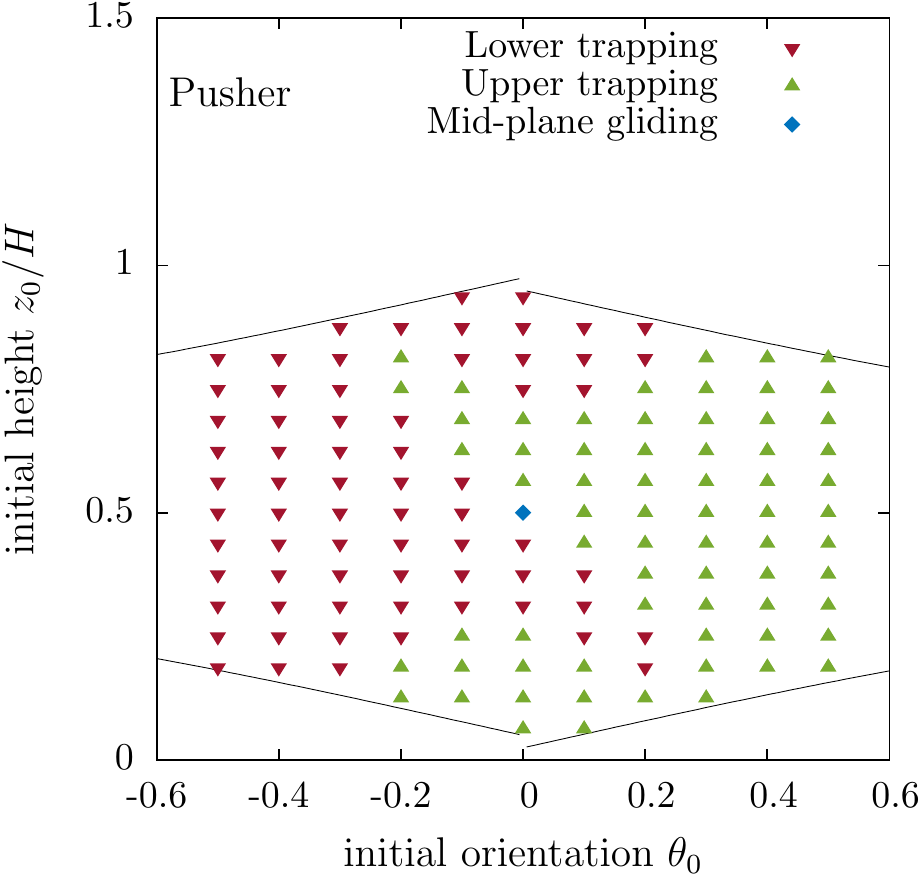}
\end{center}
\caption{(Color online) State diagram of swimming behavior in a channel of height $H=4$, for a pusher-like swimmer with $a=0.1$ and radius ratios $r_2=2, r_3=1$, using the same symbols as in Fig.~\ref{Phase-diagram}.
The pusher force-dipole hydrodynamics here lead to an amplification of the oscillations seen for neutral swimmers,
which then moves the swimmer towards trapped states,
as can be seen in the exemplary trajectories in Fig.~\ref{Traj-Pull-Push-Osc}~$(a)$.
The influence of the front-aft asymmetry was tested systematically by also varying the size of the larger front bead to $r_2=1.2$, but the corresponding state diagram does not differ qualitatively from the one shown here.
Due to the front-aft asymmetry of this three-sphere swimmer, the solid lines indicating forbidden configurations here are asymmetric.  
}
\label{Phase-diagram-Push}
\end{figure}

\begin{figure}
\begin{center}
\includegraphics[scale=0.9]{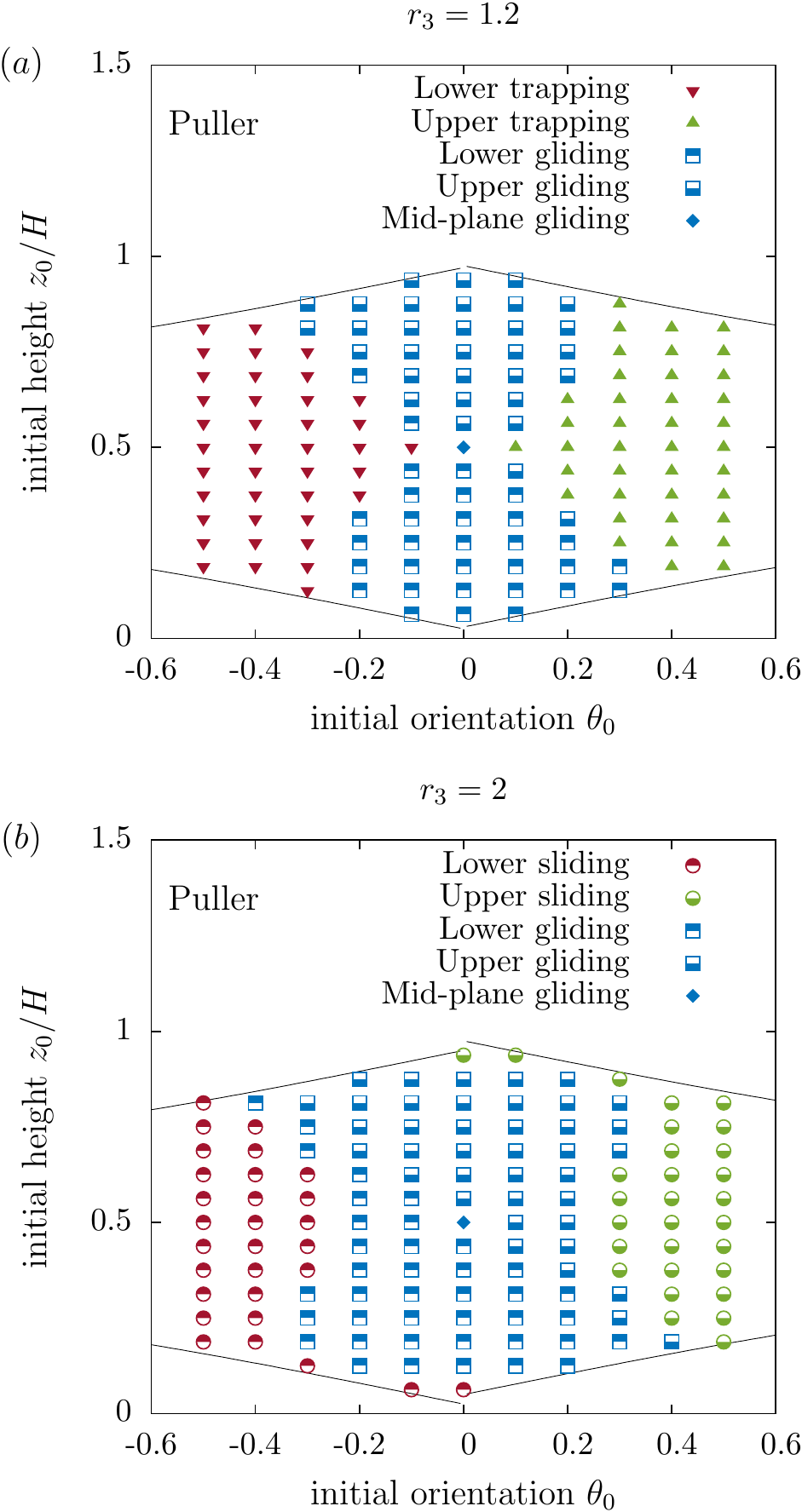}
\end{center}
\caption{(Color online) Swimming state diagram of a puller-like swimmer in a channel of height $H=4$, for $(a)$ $r_3=1.2$ and $(b)$ $r_3=2$, while the other radius ratio $r_2=1$ is held constant. 
Symbols indicate the final state of the swimmer started at the corresponding initial phase space position. 
Here half-filled (blue) boxes stand for gliding states above the corresponding wall, and half-filled circles stand for states in which the swimmer slides along one of the walls. 
The positions of the filled sides (and the corresponding colors given in the legend) then indicate which wall the respective final swimming state is nearer to.
Due to the front-aft asymmetry of the regarded three-sphere swimmers, the unaccessible phase space areas here are again asymmetric.  
$(a)$ For small~$r_3$, the swimmer either becomes trapped above one of the walls 
{or glides well above/below it.}
$(b)$ For larger $r_3$, a swimmer can either glide or start sliding along the corresponding wall, thereby maintaining a constant orientation, but is never trapped. 
}
\label{Phase-diagram-Pull}
\end{figure}

\begin{figure}
\begin{center}
\includegraphics[scale=0.85]{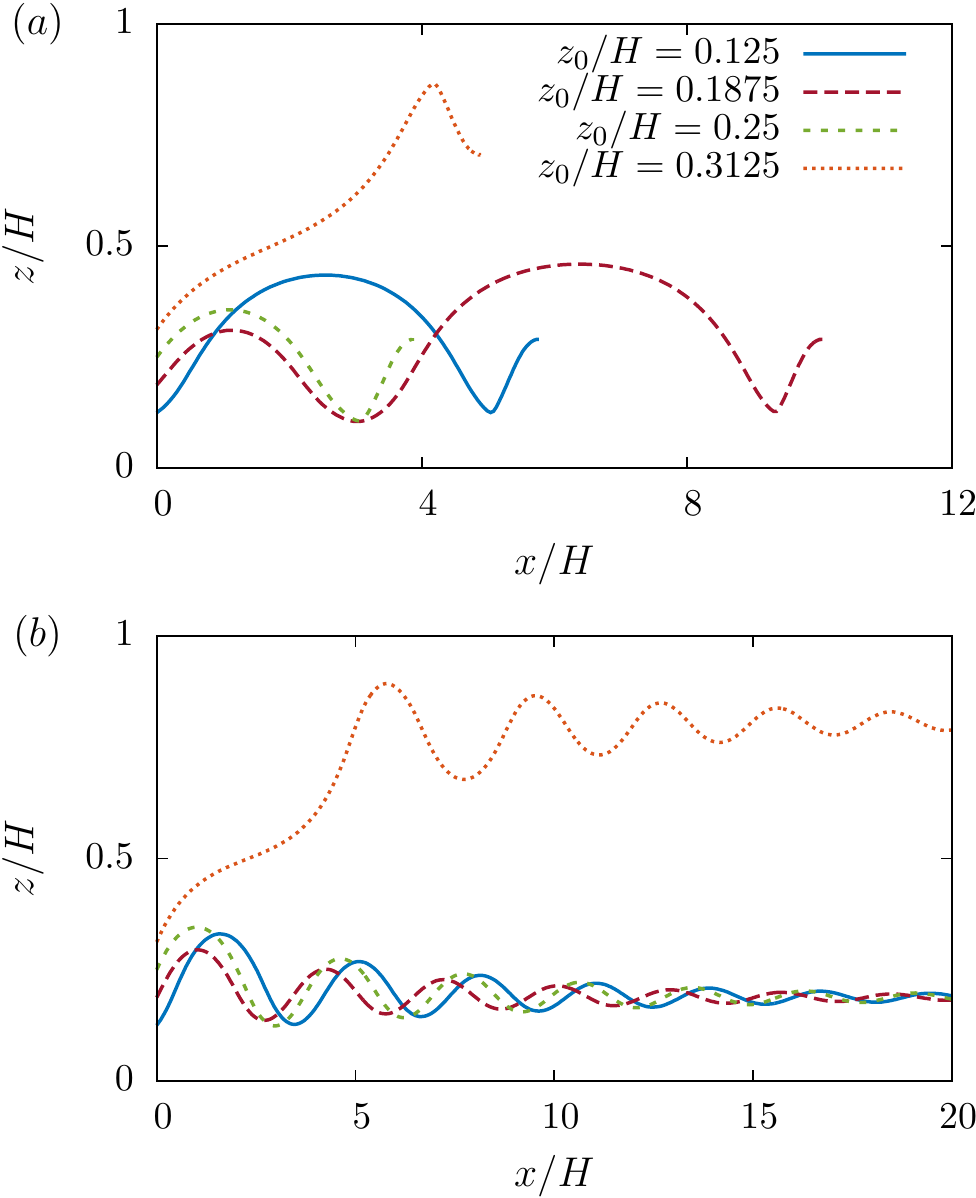}
\end{center}
\caption{
(Color online) Typical swimming trajectories of three-sphere swimmers {released at $\theta_0=0.2$} from different initial heights for $(a)$ a pusher-like swimmer with $r_2=1.2$ and $(b)$ for a puller-like swimmer with $r_3=1.2$. 
The respective other radius ratios are all set to one.
$(a)$ Pusher-like-front-heavy swimmers can no longer perform perpetual gliding motions as any oscillation is amplified until a trapped state is reached.
The end of the trajectories marks the final position in the trapped state.
$(b)$ Puller-like, aft-heavy swimmers undergo damping of their oscillations so that a straight motion parallel to the wall channels is approached in the steady state.
In both cases, the initial configuration determines which of two symmetric phase-space fixed points a swimmer will approach. 
{Here we set $a=u_0=0.1$.}
}
\label{Traj-Pull-Push-Osc}
\end{figure}


Exemplary state diagrams for a general three-sphere swimmer in a channel of a height $H=4$ are shown in Fig.~\ref{Phase-diagram-Push} for a pusher-like swimmer
and in Fig.~\ref{Phase-diagram-Pull} for a puller-like swimmer. 
For the former case, we observed one general behavior for a large range of parameters, while we found in the latter case that the behavior changes qualitatively when the radius of the enlarged sphere is increased. 
As detailed below, both these types of non-neutral swimmers show qualitative differences to the state diagram for equal-sized spheres previously discussed in Sec.~\ref{sec:stateDiagram}.

For pusher-like swimmers, the oscillatory gliding state observed for neutral swimmers is destabilized (see Fig.~\ref{Phase-diagram-Push}, where $r_2 = 2,r_3 =1 $). 
The amplitude of any initial oscillation grows rapidly with time until the swimmer ceases oscillating to reach one of two phase space fixed points which are symmetrically disposed with respect to the channel mid-height.
After transient oscillations, the swimmer reorients itself towards the nearest wall and remains in a hovering state, as can be seen in the exemplary trajectories shown in Fig.~\ref{Traj-Pull-Push-Osc}~$(a)$ 
{for various initial heights with $\theta_0=0.2$.}

Consequently, a pusher-like swimmer always ends up trapped by the channel walls with the only exception of the exactly symmetric perpetual motion along the centerline.
Depending on the initial configuration, the swimmer moves towards either the lower or the upper phase space fixed points. 
As before, the state diagram is symmetric with respect to $(z_0, \theta_0)=(H/2, 0)$, when \enquote{upper} becomes \enquote{lower} upon the corresponding point reflection and vice versa.
We have tested the qualitative robustness of this state diagram by varying the radius of the front sphere such that $r_2 = 1.2$, while keeping $r_3=1$ and have found no qualitative difference between both cases.

For puller-like swimmers, however, the behavior depends strongly on the size of the enlarged aft sphere.
Fig.~\ref{Phase-diagram-Pull}~$(a)$ shows the swimming state diagram for $r_3 = 1.2$ and $r_2=1$ resulting in a small dipolar contribution to the hydrodynamic flow field. 
In contrast to pusher-like swimmers, gliding states are here found to be generally compatible with puller hydrodynamics.
As can be seen from typical trajectories depicted in Fig~\ref{Traj-Pull-Push-Osc}~$(b)$ {for various initial heights with $\theta_0=0.2$,} the oscillations in the gliding states seem to dampen out apparently.
A strictly horizontal motion near either the upper or bottom wall is approached, termed as {lower gliding} and {upper gliding}, respectively.
As shown before for neutral swimmers, other configurations can lead to trapped states for relatively small dipolar coefficient.
However, when~$r_3$ is further increased, \eg, $r_3=2$ as shown in Fig.~\ref{Phase-diagram-Pull}~$(b)$, the non-oscillatory gliding persists, but additionally trapped states cease to exist and new {sliding} states emerge.
In these latter states, the swimmer maintains a constant non-zero orientation and undergoes a translational motion along the horizontal direction at a constant height.
{The sliding behavior emerges following a state in which the propulsive forces and the viscous forces balance each other. For a strong front-aft asymmetry, the swimmer reaches a fixed point in the comoving frame for an angle strictly less that $\pi/2$ in magnitude and undergoes a purely translational motion without oscillations parallel to the nearest wall.}

{As pointed out by de Graaf \textit{et al.}~\cite{degraaf16understanding}, the onset of the oscillatory behavior observed in neutral swimmers shown in Fig.~\ref{Oscillations} is attributed to the hydrodynamic quadrupole moment which tends to rotate the swimmer away from the nearest wall.
Analogous persistent oscillations have been observed by Zhu \textit{et al.}~\cite{zhu13} for a neutral squirmer moving in a capillary tube. 
In contrast, the dipolar contribution tends to attract a pusher toward the wall and retain a puller on the mid-channel plane.
By combining both the quadrupolar and dipolar contributions, the swimmer undergoes an oscillatory motion characterized by growing and decaying amplitudes for a pusher- and puller-type swimmer, respectively} \footnote{{See Supporting Information at [URL will be inserted by the editor] for illustrative movies showing the additional swimming states observed for puller-type swimmers with $r_2=1$.
Movie~4 shows the lower gliding state $(z_0/H=0.125, \theta_0=0.2)$ shown in Fig.~8$(b)$ for $r_3=1.2$ (solid blue line).
Movie~5 shows the upper gliding state $(z_0/H=0.3125, \theta_0=0.2)$ shown in Fig.~8$(b)$ for $r_3=1.2$ (dotted orange line).
Movie~6 illustrates the lower sliding  $(z_0/H=0.5, \theta_0=-0.3)$ for $r_3=2$.
Movie~7 illustrates the upper sliding $(z_0/H=0.5, \theta_0=0.3)$ for $r_3=2$.
For illustrative purposes, the sizes of the spheres are not shown in real scale.}}.


\subsection{Swimming stability in the mid-plane}

In the previous section, we have shown that pusher-type swimmers are trapped at the walls while puller-type swimmers undergo a gliding or sliding motion along the channel after a rapid decay of their oscillations.
An oscillatory gliding of non-varying amplitude at a constant mean height is displayed by neutral three-sphere swimmers.
For symmetry considerations, however, all three types  undergo a trivial gliding motion along the channel centerline for $z_0=H/2$ and $\theta_0=0$.

We now address the question of whether or not swimming on the channel centerline is a stable dynamical state. 
In order to proceed analytically, we restrict ourselves to the neutral swimmer case and assume for simplicity a zero initial orientation of the swimmer relative to the horizontal direction.
By combining Eqs.~\eqref{diffZdiffT} and \eqref{diffThediffT}, eliminating the time variable and integrating both sides of the resulting equation, the orientation of the swimmer is related to the distance along the channel via 
\begin{equation}
   \theta^2 = \theta_0^2 + Q(z, z_0) \, , \label{theta2_Analytics}
\end{equation}
where the integral function $Q(z, z_0)$ is given by
\begin{equation}
	Q(z, z_0) = \int_{z_0}^z \frac{2K \left( C(u) - C(H-u) \right)}
	    {V_0 + K \left( B(u) + B(H-u) \right)} \, \Intd u \, . \label{Q_neutral}
\end{equation}

\begin{figure}
	\begin{center}
	\includegraphics[scale=0.75]{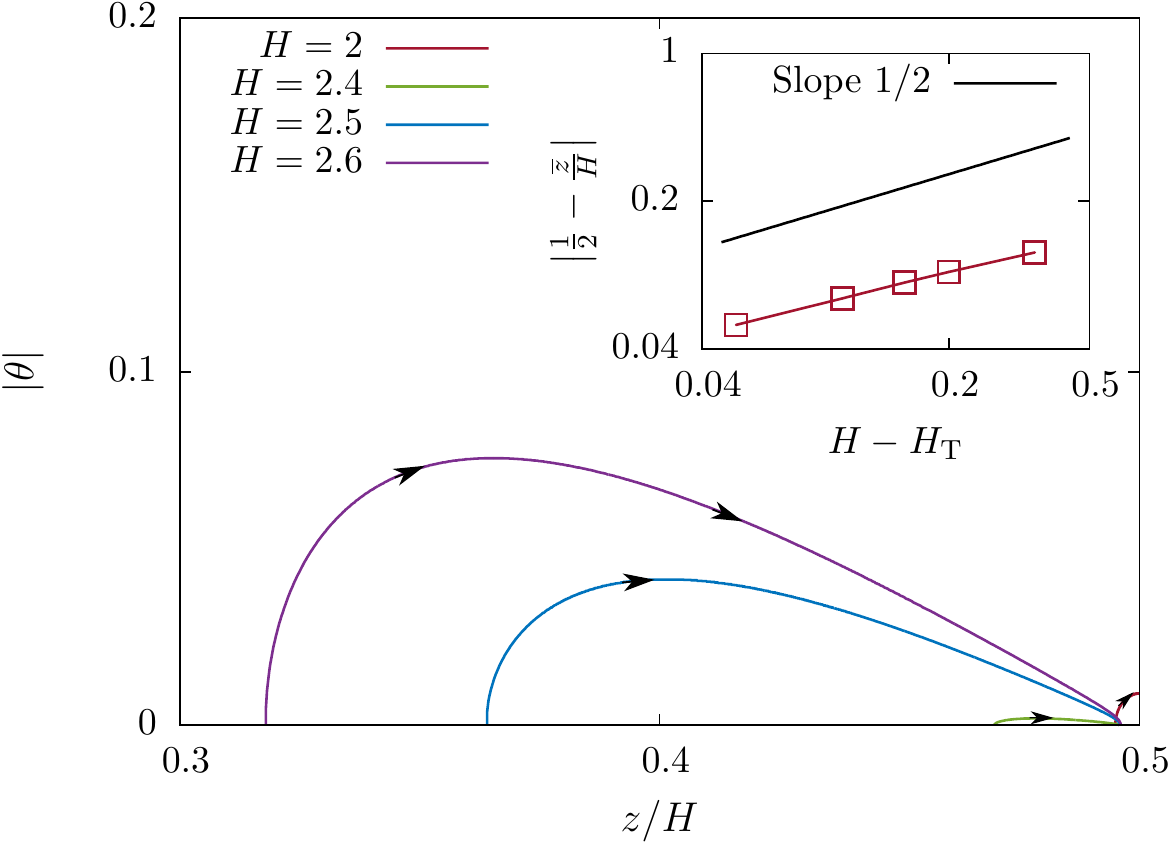}
	\end{center}
	\caption{(Color online) Trajectory of a neutral three-sphere swimmer in phase space, derived from Eq.~\eqref{theta2_Analytics} for $\theta_0=0$ and $z_0=H/2-\epsilon$ where $\epsilon=0.01$. 
		The inset shows a log-log plot of the scaled mean height relative to the channel centerline at the transition point.
		{Arrow heads show the clockwise trajectories of the swimmer in the upper phase-space.}}
	\label{PhaseSpacePlot}
\end{figure}

By evaluating the integral in Eq.~\eqref{Q_neutral} numerically and substituting the result into Eq.~\eqref{theta2_Analytics}, we obtain trajectories in the $(\theta, z)$ phase space as plotted in Fig.~\ref{PhaseSpacePlot} for $z_0 = H/2-\epsilon$ where~$\epsilon$ is an arbitrary small distance taken here as 0.01.
As expected from the state diagram shown in Fig.~\ref{Phase-diagram}, the trajectory for $H=2$ corresponds to a limit cycle around the point with $z/H=1/2$ and $\theta =0$, indicating the central oscillatory gliding motion of the swimmer.
In contrast, the trajectories are not centered at $z/H=1/2$ anymore if values of $H$ are larger than a transition value of about $H_\mathrm{T} \simeq 2.4$.

It is appropriate to denote by $\bar{z}$ the average value of the two points intersecting with the horizontal axis $z/H$. 
Around the channel centerline, the integrand on the right hand side of Eq.~\eqref{Q_neutral} can be Taylor-expanded around $z=H/2$.
Integrating the resulting equation between $H/2$ and $H/2 \pm \lambda$ yields
\begin{equation}
\theta^2 = c_2 \lambda^2 + c_4 \lambda^4 + \bigO(\lambda^6) \,  ,
\end{equation}
where $c_2$ and $c_4$ are functions of $H$ such that $c_4<0$ and $c_2$ changes sign from negative to positive as the channel height~$H$ increases beyond the transition height $H_\mathrm{T}$.
For $H>H_\mathrm{T}$, it undergoes an oscillatory motion around a mean height 
\begin{equation}
\overline{z} = \frac{H}{2} \pm \frac{1}{2} \sqrt{-\frac{c_2}{c_4}} \, . 
\end{equation}
The scaling exponent of the scaled mean height relative to the channel centerline about the transition point is readily calculated from the logarithmic derivative,
\begin{equation}
\frac{\Intd \ln \left| \frac{\overline{z}}{H}-\frac{1}{2} \right|}{\Intd \ln \left( H-H_\mathrm{T} \right)} 
=\frac{1}{2} \frac{\Intd \ln \left( \frac{1}{H} \sqrt{-\frac{c_2}{c_4}} \right)}{\Intd \ln \left( H-H_\mathrm{T} \right)}
\xrightarrow[]{H \to H_\mathrm{T}} \frac{1}{2} \, .
\end{equation}

Thus, the bifurcation is of a supercritical pitchfork-type.
In the inset of Fig.~\ref{PhaseSpacePlot}, we show the evolution of $\left| \frac{1}{2}-\frac{\bar{z}}{H}\right|$ as a function of $H-H_T$ to verify the scaling behavior derived above around the transition height.
An agreement between the theoretical value $1/2$ and the numerical results is clearly manifested.

If the channel height $H$ is further increased, the curve in Fig.~\ref{PhaseSpacePlot} will finally intersect with the line $z=0$, indicating the trapping of the swimmer. 
Such a behavior is in accord with the emergence of upper/lower trapping scenarios just above/below the central point corresponding to the central gliding motion in Fig.~\ref{Phase-diagram}~$(b)$.
Nevertheless, we note that a quantitative analysis is not available as the inclination angle may be large in this case, a situation that is beyond the simplified analytical theory proposed here.

\begin{figure}
\begin{center}
\includegraphics[scale=0.7]{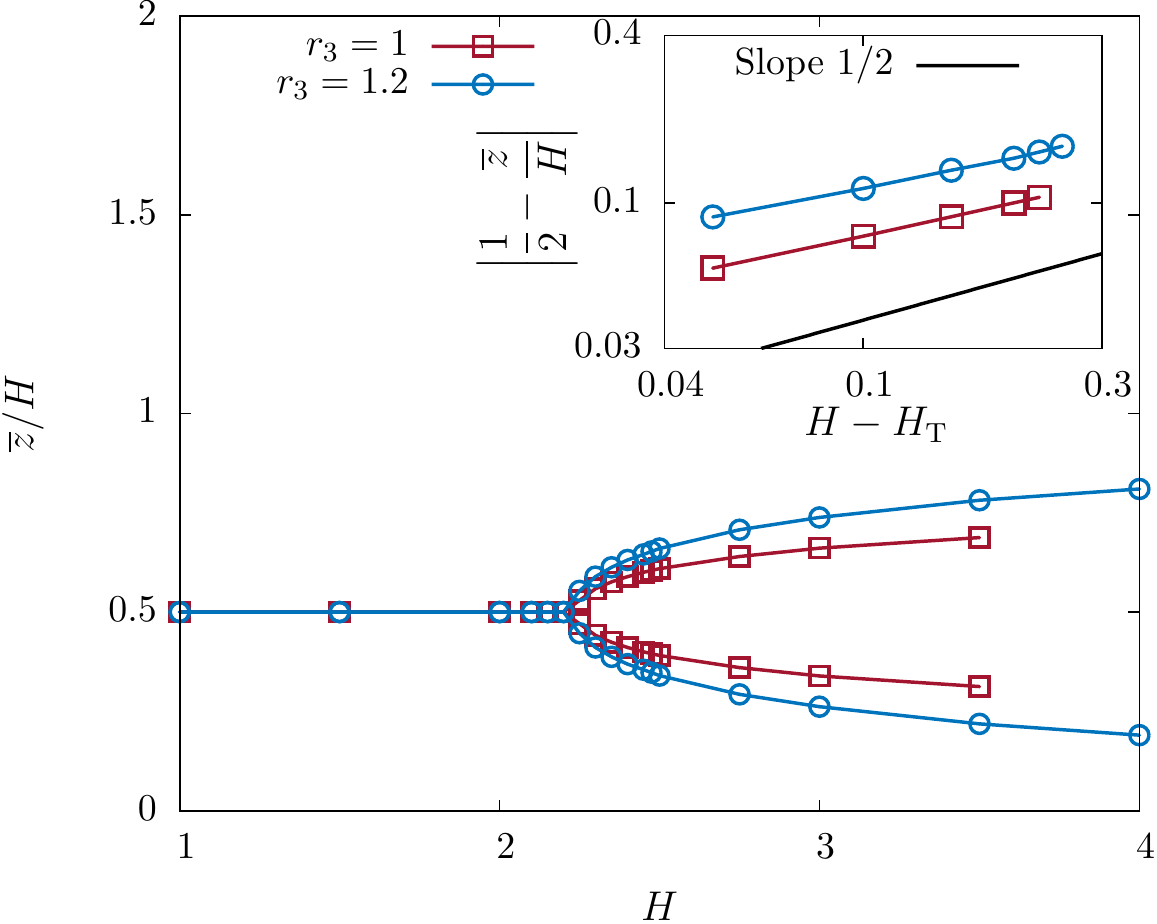}
\end{center}
\caption{(Color online) The scaled mean vertical position versus the channel height $H$ for a neutral $(r_3=1)$ and puller-type swimmer $(r_3=1.2)$.
The system undergoes a supercritical pitchfork bifurcation at $H_\mathrm{T} = 2.25$. 
Here $r_2=1$ and the swimmer is initially released from $\theta_0=0$ and $z_0=H/2 \pm \epsilon$ where $\epsilon=0.01$.
Inset: Log-log plot of the mean oscillation height (for $r_{3}=1$) and steady gliding height (for $r_{3}=1.2$) around the transition point.}
\label{Bifurkationsanalyse}
\end{figure}

We further elucidate the validity and reliability of our prediction by direct comparison with the numerical solution for a neutral {swimmer} as well as for a puller-type swimmer.
As above, we also extract $\bar{z}$ values and observe a bifurcation behavior near $H \simeq 2.25$, as clearly shown in Fig.~\ref{Bifurkationsanalyse}.
For a puller-type swimmer, $\bar{z}$ denotes the final height reached by the swimmer after the decay of oscillations.
Clearly, the bifurcation is of a pitchfork-like type as swimming in the mid-channel in the $H<H_T$ regime (c.f.~Fig.~\ref{Phase-diagram}~$(a)$ becomes unstable for $H>H_T$.
This corresponds to the appearance of the isolated points of central gliding at $z_0 =0$ and $\theta_0 =0$ in Fig.~\ref{Phase-diagram}~$(b)$ and in Fig.~\ref{Phase-diagram-Pull}~$(a)$.
Instead, two new limit cycles emerge for a neutral swimmer, indicating the lower/upper oscillatory gliding modes of motion.
For a puller swimmer, however, these new states represent stable fixed points since the oscillations are damped out in the steady limit.
We also note that, in the case of the three-sphere swimmer of equal sphere radii with $H > 3.5$, a small perturbation leads to the trapping dynamics of the swimmer, as already discussed above.
{For the value of the exponent,} however, a slight deviation from the theoretical prediction is observed (see the inset). 
We presume that this is most probably due to the approximations involved in our simplistic analytical theory.


\section{Conclusions}\label{sec:conclusions}

The dynamics of microswimmers in confined geometries reveals qualitatively new behavior due to the anisotropic nature of hydrodynamic interactions with boundaries. In this work, we characterize the motion of a swimmer in the parallel-wall channel geometry, relevant to microfluidic and Hele-Shaw cell geometries. As a model swimmer, we choose the well known three-sphere model by Najafi and Golestanian\cite{najafi04}. By considering spheres of different radii, we are able to explore the relation between the flow signature of the swimmer (pusher, puller, or neutral {swimmer}) and the observed behavior. For each type we determine the phase diagram of possible final states as a function of the initial position and orientation of the swimmer in the channel. To account for the hydrodynamic interactions with the walls, we use the method of reflections~\cite{mathijssen16jfm,degraaf16understanding}, which leads to good-quality approximations of the near-wall self- and pair mobility for spheres. 

In accord with the previously analyzed dynamics of the model swimmer close to a single planar no-slip boundary\cite{daddi18}, for a neutral swimmer (corresponding to the classical design with three identical spheres) we observe three distinct types of behavior, namely trapping at the wall, escape from the wall, and gliding at a specific distance separated from the wall, determined by the size of the swimmer and in relation to the channel width. Here, we find that the gliding state can occur both in the central area of the channel and closer to one of the walls. We then characterize the differences between puller- and pusher-type swimmer. For pusher-like swimmers, the oscillatory gliding state is unstable, and the evolution involves transient oscillations of growing amplitude, finally crossing into trapping in a hovering state at one of the walls. This observation within our numerical tests seems to be robust with respect to the changing properties of the swimmer. Puller-like swimmers, in contrast to that, exhibit a strong dependence of their modes of motion on their geometric characteristics. We find persistent gliding states compatible with the general puller hydrodynamics, with initial oscillations apparently dying out in favor of a steady solution at a fixed swimmer-to-wall distance. As the parameters of the swimmer are varied, the trapping states can vanish and sliding states appear, in which the swimmers translate at a constant height with a fixed orientation. We have also investigated analytically the stability of swimming along the centerline of the channel by considering small perturbations around the symmetric state. We find that above a critical channel width there is a pitchfork bifurcation for the oscillatory motion to appear, and we characterized it analytically. 

We believe that our findings can be useful for the design and understanding of the motion of swimming microrobots in confined geometries. Relating the initial position in the channel to the final dynamical states is particularly important for engineering microfluidic devices to sort or accumulate swimmers. The presence of boundaries leads to a variety of complex behaviors emerging for the swimmers. Our work demonstrates, however, that simple analytical approximations can still be profitably used to characterize the dynamics in many cases.

\begin{acknowledgments}
The authors gratefully acknowledge support from the DFG (Deutsche Forschungsgemeinschaft) through the projects LO~418/17-2, ME~3571/2-2, and DA~2107/1-1.
This work has been supported by the Ministry of Science and Higher Education of Poland via the Mobility Plus Fellowship awarded to ML. 
ML acknowledges funding from the Foundation for Polish Science within the START programme. 
The work of AJTMM was supported by a cross-disciplinary fellowship from the Human Frontier Science Program Organization (HFSPO - LT001670/2017).
SG gratefully acknowledges funding from the Alexander von Humboldt Foundation.
JB would like to acknowledge financial support from NSF Grant CBET
1603627.
This article is based upon work from COST Action MP1305, supported by COST (European Cooperation in Science and Technology).
\end{acknowledgments}


\appendix

\section*{Appendices}

In Appendix~\ref{appendix:greensFcts}, we derive the Green's functions in a channel between two no-slip walls using a two-dimensional Fourier transform technique.
We then describe in Appendix~\ref{appendix:Arnold} the method of reflections and express the Green's functions in the channel in terms of an infinite series of images.
In Appendix~\ref{appendix:3sphereSwimmer}, we provide an overview on the dynamics of a general three-sphere swimmer in an unbounded fluid domain and show that the induced far-flow field can conveniently be described by a combination of dipolar and quadrupolar flows.
Further mathematical details are contained in Appendix~\ref{appendix:formulas}.

\allowdisplaybreaks

\section{Green's functions}\label{appendix:greensFcts}

In this appendix, we use a two-dimensional Fourier transform technique to derive the Green's functions in a channel between two no-slip walls.
The solution method consists of reducing the partial differential equations  \eqref{stokesEq} into ordinary differential equations in the direction perpendicular to the walls, whereas the spatial dependence of the hydrodynamic fields in the plane parallel to the wall are Fourier transformed into the wavenumber domain.
Upon inverse Fourier transformation, the Green's functions can conveniently be expressed in terms of Bessel integrals of the first kind.

We define the two-dimensional Fourier transform
\begin{equation}
\mathscr{F} \{ f(\bRho) \} =: \tilde{f}(\bQ) = 
\int_{\mathbb{R}^2} f(\bRho) e^{-i \bQ . \bRho} \, \Intd \bRho \, , 
\label{2DFourierTransform}
\end{equation}
together with its inverse transform
\begin{equation}
\mathscr{F} ^{-1} \{ \tilde{f} (\bQ) \} =: f(\bRho) =
\frac{1}{(2 \pi)^2}
\int_{\mathbb{R}^2} \tilde{f} (\bQ) e^{i \bQ . \bRho} \, \Intd \bQ \, , 
\label{inverse2DFourierTransform}
\end{equation}
where  $\bRho = (x,y)$ is the projection of the vector~$\vect{r}$ onto the plane $z=0$, and $\bQ = (q_x, q_y)$ sets the coordinates in Fourier space.

It is more convenient to make use of the orthogonal basis introduced previously by Bickel~\cite{bickel07, bickel06}, in which the velocity vector field is decomposed into transverse, longitudinal, and normal components.
Accordingly, the Fourier-transformed components of the velocity field in the Cartesian coordinate basis $\tilde{v}_x$ and $\tilde{v}_y$ are related to the longitudinal and transverse components in the new basis $\tilde{v}_l$ and $\tilde{v}_t$ via the orthogonal transformation
\begin{equation}  \label{transformation}
\left( 
      \begin{array}{c}
      \tilde{v}_x \\
      \tilde{v}_y 
      \end{array}
\right)
=
\frac{1}{q}
\left( 
      \begin{array}{cc}
      q_x & q_y\\
      q_y & -q_x
      \end{array}
\right)
\left( 
      \begin{array}{c}
      \tilde{v}_l \\
      \tilde{v}_t 
      \end{array}
\right) \, ,
\end{equation}
wherein $q := |\vect{q}|$ is the wavenumber.
The longitudinal and transverse components of the force $f_l$ and $f_t$ follow forthwith using an analogous transformation matrix.

We now assume that the point force is acting inside the channel at location $\R_0=(0,0,h)$, where $0<h<H$.	
Upon two-dimensional Fourier transform, Eqs.~\eqref{stokesEq} governing the fluid motion yield ordinary differential equations in the variable $z$.
Specifically~\cite{daddi18epje}
\begin{align}
\eta (-q^2 \vt_l+ \vt_{l,zz}) - iq \tilde{p} +f_l \, \delta (z-h) &= 0 \, , \label{longitudinalStokesGleischung} \\
\eta (-q^2 \vt_t + \vt_{t,zz})  +f_t \, \delta (z-h) &= 0 \, , \label{transverseStokesGleischung} \\
\eta (-q^2 \vt_z+ \vt_{z,zz}) - \tilde{p}_{,z} + f_z \,  \delta (z - h) &= 0 \, , \label{axialStokesGleischung} \\
iq \vt_l+ \vt_{z,z} &= 0 \, , \label{incompressibilitaetsGleischung}
\end{align}
where a comma in a subscript stands for a partial derivative. 

The velocity transverse component $\vt_t$ can directly be obtained by solving Eq.~\eqref{transverseStokesGleischung}.
By combining Eqs.~\eqref{axialStokesGleischung} and \eqref{longitudinalStokesGleischung}, the pressure field can readily be eliminated.
As the continuity equation \eqref{incompressibilitaetsGleischung} provides a direct relation between the longitudinal and normal components,
a fourth-order ordinary differential equation for $\vt_z$ is obtained, namely~\cite{bickel07}
\begin{equation}
\vt_{z,zzzz} - 2q^2 \vt_{z,zz} + q^4 \vt_z = \frac{q^2}{\eta} \, f_z \, \delta (z - h) + \frac{iq}{\eta}  \, f_l \, \delta'(z-h) \, ,
\label{fourthOrderDifferentialEqn}
\end{equation}
wherein $\delta'$ is the derivative of the Dirac delta function.

The Green's functions in 2D Fourier space can thus be identified from
\begin{equation}
\left( 
      \begin{array}{c}
      \vt_t\\
      \vt_l\\
      \vt_z
      \end{array}
\right)
=
\left( 
      \begin{array}{ccc}
      \tilde{\G}_{tt} & 0 & 0\\
      0 & \tilde{\G}_{ll} & \tilde{\G}_{lz} \\
      0 & \tilde{\G}_{zl} & \tilde{\G}_{zz} 
      \end{array}
\right)
\left( 
      \begin{array}{c}
      f_t \\
      f_l \\
      f_z
      \end{array}
\right) \, .
\label{greenFunctions}
\end{equation}

In the following, we present an analytical solution for the fluid velocity field in the channel by considering the solutions for the transverse and normal components independently.

\subsection{Transverse velocity}

The general solution of Eq.~\eqref{transverseStokesGleischung} inside a channel of width~$H$ can be written as
\begin{equation}
	\vt_t = A_1 e^{qz} + B_1 e^{-qz} \, ,
\end{equation}
for $0 \le z \le h$, and
\begin{equation}
	\vt_t = A_2 e^{q(H-z)} + B_2 e^{-q(H-z)} \, ,
\end{equation}		 
for $h \le z \le H$, wherein $A_\alpha$ and $B_\alpha$, for $\alpha \in \{1,2\}$, are wavenumber-dependent quantities to be determined from the underlying boundary conditions.
The no-slip condition at the walls yields $\vt_t(z=0) = \vt_t(z=H)=0$.
Additionally, the Dirac delta function implies the discontinuity of the first derivative at the point-force position.
Specifically
\begin{equation}
	\left. \vt_{t,z} \right|_{z=h^+} - \left. \vt_{t,z} \right|_{z=h^-} = -\frac{f_t}{\eta} \, ,
\end{equation}
by requiring the natural continuity of the transverse velocity at $z=h$.

Solving for the four unknown quantities yields
\begin{align}
	A_1 &= \frac{f_t}{2q\eta} \frac{\sinh \left(q(H-h)\right)}{\sinh (qH)} \, , \\
	A_2 &= \frac{f_t}{2q\eta} \frac{\sinh (qh)}{\sinh (qH)} \, , 
\end{align}
with $B_1 = -A_1$ and $B_2 =-A_2$.

\subsection{Normal velocity}

The general solution of Eq.~\eqref{fourthOrderDifferentialEqn} for the normal velocity is given by
\begin{equation}
	\vt_z = (C_1+D_1 z)e^{qz} + (E_1+F_1 z)e^{-qz} \, , 
\end{equation}	
for $0 \le z \le h$, and
\begin{equation}	
	\vt_z = \big( C_2+D_2(H-z) \big) e^{q(H-z)} 
		  + \big( E_2+F_2(H-z) \big) e^{-q(H-z)} \, , 
\end{equation}
for $h \le z \le H$.
Here $C_\alpha, D_\alpha, E_\alpha$, and $F_\alpha$, $\alpha \in \{1,2\}$, are unknown wavenumber-dependent functions to be determined from the boundary conditions.	
The no-slip condition at the channel walls yields $\vt_z(z=0)=\vt_z(z=H)=0$.
In addition, since 
\begin{equation}
	\vt_l= \frac{i}{q} \, \vt_{z,z} \, , \label{longitudinalVonNormal}
\end{equation}
as can be inferred from the continuity equation \eqref{incompressibilitaetsGleischung}, we further require that $\vt_{z,z}(z=0)=\vt_{z,z}(z=H)=0$.
Moreover, the Dirac delta function implies the discontinuity of the third derivative of the normal velocity, 
\begin{equation}
	\left. \vt_{z,zzz} \right|_{z=h^+} - \left. \vt_{z,zzz} \right|_{z=h^-} =
	\frac{q^2 f_z}{\eta} \, ,
\end{equation}
while the derivative of the delta function implies the discontinuity of the second derivative, 
\begin{equation}
	\left. \vt_{z,zz} \right|_{z=h^+} - \left. \vt_{z,zz} \right|_{z=h^-} =
	\frac{iq f_l}{\eta} \, .
\end{equation}

By requiring the continuity of the normal and longitudinal velocities at the point-force position, making use of \eqref{longitudinalVonNormal}, and solving the resulting system of eight equations for the unknown quantities, we readily obtain
\begin{align}
  C_1&= \frac{i f_l}{8\eta q b_0} \Big( S_-(-u, -U) - S_+(u, U) \Big) \notag \\
  &- \frac{f_z}{8\eta q b_0} \Big( S_-(u, U) + S_+(-u, -U) \Big) \, , \notag \\
  D_1&= \frac{1}{ 4\eta b_0} \Big(  S_-(u,U) f_z + i S_+ (u,U) f_l \Big) \, , \notag \\
  E_1&= -C_1  \, , \notag \\
  F_1&= \frac{1}{ 4\eta b_0} \Big(  S_+ (-u,-U)f_z - i  S_- (-u,-U) f_l \Big) \, , \notag 
 \end{align}
where we have defined the dimensionless quantities
\begin{equation}
	u = qh \, , \quad U = qH \, , \quad b_0 =  2+4 U^2 - 2 \cosh(2 U)  \, , \notag
\end{equation}
in addition to
\begin{equation}
	S_\pm(x_1,x_2) = b_1 (x_1,x_2) \pm b_2 (x_1,x_2) \, , 
\end{equation}
where
\begin{align}
  b_1 (x_1,x_2) &= 2 \big( \cosh(x_1-2 x_2) -  \cosh(x_1) \notag \\
                &+ 2U(u-U) \exp(-x_1) \big) \, , \notag \\
  b_2 (x_1,x_2) &= 4 \left( U \sinh(x_1) - u \exp(x_1-x_2) \sinh x_2 \right) \, . \notag 
\end{align}

The wavenumber-dependent functions for the fluid domain $h \le z < H$ are obtained as
\begin{equation}
	C_2 = -C_1 |_{h \to H-h} \, , 
\end{equation}
and analogously for $D_2$, $E_2$, and $F_2$.

Upon inverse Fourier transformation, the Green's functions can conveniently be written in terms of convergent infinite integrals over the wavenumber~$q$, as~\cite{daddi16}, 
\begin{subequations}
   \begin{align}
    \G_{xx} (\R, z_0) =& \frac{1}{4\pi}  \int_0^{\infty} \bigg( \Gt_{+} (q,z,z_0) J_0 (\rho q)  \notag \\
      &+   \Gt_{-} (q,z,z_0) J_2 (\rho q) \cos (2\theta) \bigg) q \, \Intd q \, ,  \label{Gxx} \\
    \G_{yy} (\R, z_0) =& \frac{1}{4\pi}  \int_0^{\infty} \bigg( \Gt_{+} (q,z,z_0) J_0 (\rho q)  \notag \\
      &- \Gt_{-} (q,z,z_0) J_2 (\rho q) \cos (2\theta) \bigg) q \, \Intd q \, ,  \label{Gyy} \\
    {\G}_{zz} (\R, z_0) =& \frac{1}{2\pi}
    \int_{0}^{\infty}  \Gt_{zz} (q,z,z_0) J_0 (\rho q) q \, \Intd q \, , \label{Gzz} 
    \end{align}
    \label{greenFunctionsBackToRealSpace}
\end{subequations}
for the diagonal components, and 
\begin{subequations}
	\begin{align}
		\G_{xy} (\R, z_0) =& \frac{\sin (2\theta) }{4\pi} \int_0^\infty \Gt_{-} (q,z,z_0) J_2 (\rho q) q \, \Intd q \, , \label{Gxy} \\
		    \G_{rz} (\R, z_0) =& \frac{i}{2\pi} \int_{0}^{\infty} \Gt_{lz} (q,z,z_0) J_1 (\rho q) q \, \Intd q \, , \label{Grz}\\
		    \G_{zr} (\R, z_0) =& \frac{i}{2\pi} \int_{0}^{\infty} \Gt_{zl} (q,z,z_0) J_1 (\rho q) q \, \Intd q \, , \label{Gzr}
	\end{align}
\end{subequations}
for the off-diagonal components.
Here $\rho^2 := {x^2 + y^2}$ and $\theta := \arctan (y/x)$ is the polar angle.
In addition, $J_n$ denotes the Bessel function~\cite{abramowitz72} of the first kind of order~$n$.
Moreover, 
\begin{equation}
 \Gt_{\pm} (q,z) := \Gt_{tt}(q,z) \pm \Gt_{ll}(q,z) \, . \nonumber
\end{equation}

The components in Cartesian coordinates can be obtained from the usual transformation $\G_{xz}=\G_{rz}\cos\theta$, $\G_{yz}=\G_{rz}\sin\theta$, $\G_{zx}=\G_{zr}\cos\theta$, and $\G_{zy}=\G_{zr}\sin\theta$.
Moreover, note that $\G_{yx} = \G_{xy}$.


\section{Images of a Stokeslet between parallel no-slip walls}\label{appendix:Arnold}

Here we describe the flow due to a point force
in a Stokesian liquid between two parallel no-slip boundaries, in terms of an infinite series of image reflections.
This method is complementary to the one developed by Liron and Mochon~\cite{liron76}, who first gave the Green's function solution in terms of a Hankel transformation.
A detailed comparison between these two methods is given by Mathijssen \textit{et al.}~\cite{mathijssen16jfm} for Stokeslets and higher order multipoles between a no-slip wall and a free surface.
Previous studies have also used the reflection method to investigate the flow produced by mobile colloids \citep{Ozarkar2008}.

\begin{figure*}
\begin{center}
\includegraphics[width=\linewidth]{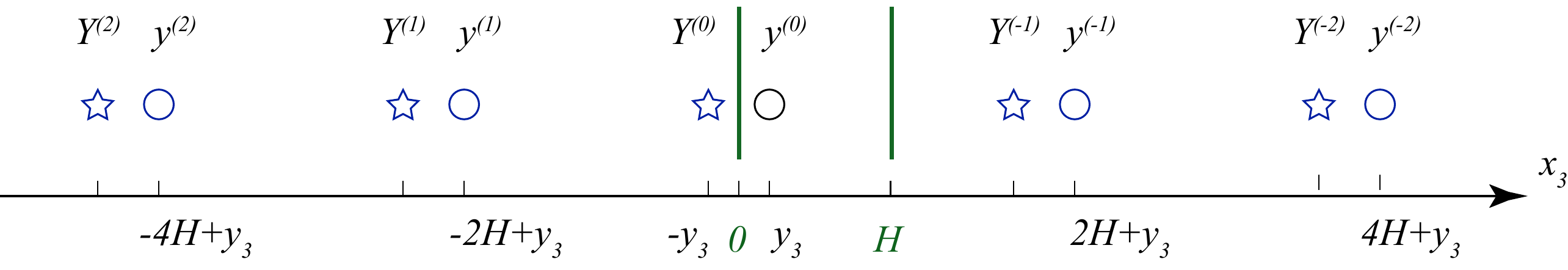} 
\end{center}
\caption{
Diagram showing the image reflections of a Stokeslet, located at $y_3$, between two parallel no-slip walls, located at $x_3 = 0,H$. The images reflected an even number of times (circles) are located at $y_3^{(m)} = y_3 - 2mH$ and those reflected an odd number of times (stars) are located at $Y_3^{(m)} = -y_3 - 2mH$.
}
\label{fig:geometryDiagram}
\end{figure*}

to connect with previous notations in Ref.~\onlinecite{mathijssen16jfm}, we rewrite the Stokes equations \eqref{stokesEq} into the form
\begin{eqnarray}
\label{eq:StokesEqnA}
\vec{\nabla}  P(\vec{x}) - \eta \nabla^2 \vec{v}(\vec{x})  &=& \vec{f} ~ \delta(\vec{x} - \vec{y}),
\\
\label{eq:StokesEqnB}
\vec{\nabla} \cdot \vec{v}(\vec{x}) &=& 0,
\end{eqnarray}
where the fluid velocity is $\vec{v}(\vec{x},t)$, the pressure field is $P(\vec{x},t)$, the fluid position is $\vec{x} = (x_1, x_2, x_3)$ at time $t$, and the point force density is $\vec{f}  ~ \delta(\vec{x} - \vec{y})$ (Stokeslet) that acts on the liquid at position $\vec{y} = (y_1, y_2, y_3=h)$.
The velocity field must satisfy the no-slip boundary condition, $\vec{v}(\vec{x}) = 0$ at the channel walls $x_3 = 0, H$.

In the absence of boundaries, the flow is given by the Oseen tensor,
\begin{eqnarray}
v_i^\textmds{S}(\vec{x},\vec{y},\vec{f}) &=& \mathcal{J}_{ij} (\vec{x},\vec{y}) f_j,
\\
\label{eq:OseenTensor}
\mathcal{J}_{ij}(\vec{x},\vec{y}) &=& \frac{1}{8\pi \eta} \left( \frac{\delta_{ij}}{r} + \frac{r_i r_j}{r^3} \right), \quad i,j \in \{1,2,3\}, 
\end{eqnarray}
where $\vec{r} = \vec{x} - \vec{y}$, $r=|\vec{r}|$, $\delta_{ij}$ is the Kronecker delta, and repeated indices are summed over.
The pressure that completes this solution is $P(\vec{x},\vec{y},\vec{f}) = \mathcal{P}_j f_j$ with $ \mathcal{P}_j = r_j / 4\pi r^3$.
We now aim to solve the flow in a channel in terms of this Oseen tensor and derivatives thereof only, using the method reflections.

On the one hand, for the case of only a single boundary being present, i.e., $H\rightarrow\infty$ in our system, Blake~\cite{blake71} first derived the Stokeslet flow in terms of an image system.
The image is located at the position $\vec{Y}^{(0)} = (y_1, y_2, -y_3) = \mbox{M} \cdot \vec{y}$, where the diagonal mirror matrix is $\mbox{M} = \mbox{diag}(1,1,-1)$.
The image tensor is found by applying the reflection operator, B, of the `Bottom' wall to the Stokeslet.
This operator B$(\lambda)$ is a function of the distance from the wall to the Stokeslet, which is $\lambda = y_3$ here.
Hence, we have
\begin{equation}
\mathcal{B}_{ij} (\vec{x},\vec{Y}^{(0)})  = \mbox{B} ~ \mathcal{J}_{ij}(\vec{x}, \vec{y}^{(0)}) \, ,
\end{equation}
where the Blake solution can then be written in terms of the Oseen tensor as
\begin{align}
\mathcal{B}_{ij} &(\vec{x},\vec{Y}^{(0)}) \notag \\
&=
(- \delta_{jk} + 2 \lambda \delta_{k3} \tilde{\partial}_j 
+ \lambda^2 \mbox{M}_{jk} \tilde{\nabla}^2) \mathcal{J}_{ik}(\vec{x}, \vec{Y}^{(0)}) 
\label{eq:blakeTensor1} \\
&=
(- \delta_{jk} + 2 y_3 \delta_{k3} \tilde{\partial}_j 
+ y_3^2 \mbox{M}_{jk} \tilde{\nabla}^2) \mathcal{J}_{ik}(\vec{x}, \vec{Y}^{(0)}) \, , \notag
\end{align}
where the derivatives $\tilde{\partial}_j = \tfrac{\partial}{\partial y_j} = \mbox{M}_{jl} \tfrac{\partial}{\partial Y_l^{(0)}}$  
and $\tilde{\nabla}^{2} = \tilde{\partial}_l \tilde{\partial}_l$ 
are with respect to the force position $\vec{y}$. 
The first row of \tbl{tab:images2} lists this tensor $\mathcal{B}_{ij} (\vec{x},\vec{Y}^{(0)})$ as the first `Bottom' reflection.
The overall flow field is then given by
\begin{eqnarray}
u^\textmds{B}_i (\vec{x}, \vec{y}) &=& \left[ \mathcal{J}_{ij}(\vec{x},\vec{y}^{(0)}) + \mathcal{B}_{ij} (\vec{x},\vec{Y}^{(0)}) \right] f_j.
\end{eqnarray}
On the other hand, if only the top wall is present at $x_3 = H$, the distance from the wall to the Stokeslet is $\lambda = y_3 - H$ and the reflection is located at $\vec{Y}^{(-1)} = (y_1, y_2, 2H - y_3)$.
The image tensor is then given by applying the reflection operator, T$(\lambda)$, of the `Top' wall to the Stokeslet,
\begin{align}
 \mathcal{T}_{ij} & (\vec{x},\vec{Y}^{(-1)}) \notag \\
 &= \mbox{T} ~ \mathcal{J}_{ij}(\vec{x}, \vec{y}^{(0)}) \notag \\
&=
(- \delta_{jk} + 2 \lambda \delta_{k3} \tilde{\partial}_j 
+ \lambda^2 \mbox{M}_{jk} \tilde{\nabla}^2) \mathcal{J}_{ik}(\vec{x}, \vec{Y}^{(-1)}) \label{eq:freeBoundaryTensor}
\\
&=
(- \delta_{jk} - 2 \{H-y_3\} \delta_{k3} \tilde{\partial}_j \notag \\
&\quad\,+ \{H-y_3\}^2 \mbox{M}_{jk} \tilde{\nabla}^2) \mathcal{J}_{ik}(\vec{x}, \vec{Y}^{(-1)}) \, . \notag 
\end{align}
The second row of \tbl{tab:images2} lists this result as the first `Top' reflection.
The overall flow, given by adding (\ref{eq:OseenTensor}) and (\ref{eq:freeBoundaryTensor}), satisfies the no-slip condition exactly on the top surface.

Next, when there are two parallel plates, one can continue using the method of images by computing the reflections of the reflections, and again the reflections of those, in order to generate an infinite series of images. Each image system consists of Stokeslets and derivatives thereof, thus satisfy the Stokes equations, and by adding more reflections the boundary conditions on both surfaces will by satisfied asymptotically.
We first determine the positions of the image systems,
\begin{eqnarray}
\label{eq:imagePositions1}
\vec{y}^{(m)} &=& (y_1, y_2, y_3 - 2mH), \quad m = 0, \pm1, \pm2, \dots, \quad
\\
\label{eq:imagePositions2}
\vec{Y}^{(m)} &=& (y_1, y_2, - y_3 - 2mH), \quad m = 0, \pm1, \pm2, \dots, \quad
\end{eqnarray}
where the images at $\vec{Y}^{(m)}$ are reflected an odd number of times and the images at $\vec{y}^{(m)}$ an even number of times. 
The original Stokeslet is also included here at position $\vec{y} = \vec{y}^{(0)}$.
The resulting series of images is shown in \fig{fig:geometryDiagram}. 
Then we must determine the functional form of the image tensors, $\mathcal{G}_{ij}(\vec{x}, \vec{y}^{(m)})$ and $\mathcal{G}_{ij}(\vec{x}, \vec{Y}^{(m)})$. 
For a given image, this is done by replacing all the Oseen tensors $\mathcal{J}_{ij}$ in the previous image system by the appropriate Blake tensor.
The key idea is that the newly obtained reflection is again an expression in terms of Oseen tensors, and derivatives thereof, which can then be replaced again for the next reflection.

\begin{table*}
\begin{footnotesize}
\begin{tabular}{|l||l|l|l|}
\hline
(n) 	& Position		& Replace 									& with 
\\ \hline
(0) 	& $\vec{y}^{(0)}$	&  ---  	& $\mathcal{J}_{ij}(\vec{x}, \vec{y}^{(0)})$     \\ 
(1) 	& $\vec{Y}^{(0)}$	&  $\mbox{B} \mathcal{J}_{ij}(\vec{x}, \vec{y}^{(0)})$  	& $(- \delta_{jk} + 2 y_3 \delta_{k3} \tilde{\partial}_j + y_3^2 \mbox{M}_{jk} \tilde{\nabla}^2) \mathcal{J}_{ik}(\vec{x}, \vec{Y}^{(0)})$     \\ 
(2) 	& $\vec{Y}^{(-1)}$	&  $\mbox{T} \mathcal{J}_{ij}(\vec{x}, \vec{y}^{(0)})$  	& $(- \delta_{jk} - 2 (H-y_3) \delta_{k3} \tilde{\partial}_j + (H-y_3)^2 \mbox{M}_{jk} \tilde{\nabla}^2) \mathcal{J}_{ik}(\vec{x}, \vec{Y}^{(-1)})$     \\  
(3) 	& $\vec{y}^{(-1)}$	&  $\mbox{T} \mathcal{J}_{ij}(\vec{x}, \vec{Y}^{(0)})$  	& $ (- \delta_{jk} - 2 (H+y_3) \delta_{k3} \mbox{M}_{jl} \tilde{\partial}_l + (H+y_3)^2 \mbox{M}_{jk} \tilde{\nabla}^2) \mathcal{J}_{ik}(\vec{x}, \vec{y}^{(-1)})$    \\ 
(4) 	& $\vec{y}^{(1)}$	& $\mbox{B} \mathcal{J}_{ij}(\vec{x}, \vec{Y}^{(-1)})$  	& $(- \delta_{jk} + 2 (2H-y_3) \delta_{k3} \mbox{M}_{jl} \tilde{\partial}_l + (2H-y_3)^2 \mbox{M}_{jk} \tilde{\nabla}^2) \mathcal{J}_{ik}(\vec{x}, \vec{y}^{(1)})$     \\ 
(5) 	& $\vec{Y}^{(1)}$	& $\mbox{B} \mathcal{J}_{ij}(\vec{x}, \vec{y}^{(-1)})$  	& $(- \delta_{jk} + 2 (2H+y_3) \delta_{k3} \tilde{\partial}_j + (2H+y_3)^2 \mbox{M}_{jk} \tilde{\nabla}^2) \mathcal{J}_{ik}(\vec{x}, \vec{Y}^{(1)})$     \\ 
(6) 	& $\vec{Y}^{(-2)}$	& $\mbox{T} \mathcal{J}_{ij}(\vec{x}, \vec{y}^{(1)})$  	& $ (- \delta_{jk} - 2 (3H-y_3) \delta_{k3} \tilde{\partial}_j + (3H-y_3)^2 \mbox{M}_{jk} \tilde{\nabla}^2) \mathcal{J}_{ik}(\vec{x}, \vec{Y}^{(-2)})$     \\  
(7) 	& $\vec{y}^{(-2)}$	& $\mbox{T} \mathcal{J}_{ij}(\vec{x}, \vec{Y}^{(1)})$  	& $ (- \delta_{jk} - 2 (3H+y_3) \delta_{k3} \mbox{M}_{jl} \tilde{\partial}_l + (3H+y_3)^2 \mbox{M}_{jk} \tilde{\nabla}^2) \mathcal{J}_{ik}(\vec{x}, \vec{y}^{(-2)})$    \\ 
(8) 	& $\vec{y}^{(2)}$	& $\mbox{B} \mathcal{J}_{ij}(\vec{x}, \vec{Y}^{(-2)})$  	& $(- \delta_{jk} + 2 (4H-y_3) \delta_{k3} \mbox{M}_{jl} \tilde{\partial}_l + (4H-y_3)^2 \mbox{M}_{jk} \tilde{\nabla}^2) \mathcal{J}_{ik}(\vec{x}, \vec{y}^{(2)})$     \\ 
(9) 	& $\vec{Y}^{(2)}$	& $\mbox{B} \mathcal{J}_{ij}(\vec{x}, \vec{y}^{(-2)})$  	& $(- \delta_{jk} + 2 (4H+y_3) \delta_{k3} \tilde{\partial}_j + (4H+y_3)^2 \mbox{M}_{jk} \tilde{\nabla}^2) \mathcal{J}_{ik}(\vec{x}, \vec{Y}^{(2)})$     \\ 
(10) 	& $\vec{Y}^{(-3)}$	& $\mbox{T} \mathcal{J}_{ij}(\vec{x}, \vec{y}^{(2)})$  	& $ (- \delta_{jk} - 2 (5H-y_3) \delta_{k3} \tilde{\partial}_j + (5H-y_3)^2 \mbox{M}_{jk} \tilde{\nabla}^2) \mathcal{J}_{ik}(\vec{x}, \vec{Y}^{(-3)})$     \\  
(11) 	& $\vec{y}^{(-3)}$	& $\mbox{T} \mathcal{J}_{ij}(\vec{x}, \vec{Y}^{(2)})$  	& $ (- \delta_{jk} - 2 (5H+y_3) \delta_{k3} \mbox{M}_{jl} \tilde{\partial}_l + (5H+y_3)^2 \mbox{M}_{jk} \tilde{\nabla}^2) \mathcal{J}_{ik}(\vec{x}, \vec{y}^{(-3)})$    \\ 
(12) 	& $\vec{y}^{(3)}$	& $\mbox{B} \mathcal{J}_{ij}(\vec{x}, \vec{Y}^{(-3)})$  	& $(- \delta_{jk} + 2 (6H-y_3) \delta_{k3} \mbox{M}_{jl} \tilde{\partial}_l + (6H-y_3)^2 \mbox{M}_{jk} \tilde{\nabla}^2) \mathcal{J}_{ik}(\vec{x}, \vec{y}^{(3)})$     \\ 
(13) 	& $\vec{Y}^{(3)}$	& $\mbox{B} \mathcal{J}_{ij}(\vec{x}, \vec{y}^{(-3)})$  	& $(- \delta_{jk} + 2 (6H+y_3) \delta_{k3} \tilde{\partial}_j + (6H+y_3)^2 \mbox{M}_{jk} \tilde{\nabla}^2) \mathcal{J}_{ik}(\vec{x}, \vec{Y}^{(3)})$     \\ 
~ \vdots & ~ \vdots    & ~ \vdots       & ~ \vdots     \\ \hline
\end{tabular}
\end{footnotesize}
\caption{Recursion relations for the successive image systems of a Stokeslet between two parallel no-slip walls. 
The first image system of the Oseen tensor from reflection in the bottom wall is the Blake tensor, and the second image from reflection at the top interface is the mirrored Blake tensor. 
Subsequent image systems are obtained from further reflection operations with $\mbox{B}$ denoting the ``bottom'' wall and $\mbox{T}$ the ``top'' wall, that operate linearly on all the Oseen tensor terms $\mathcal{J}_{ij}$ of the image system tensor $\mathcal{G}_{ij}$.}
\label{tab:images2}
\end{table*}

To see this, we explicitly consider the second (T) reflection of the first (B) image (\ref{eq:blakeTensor1}).
This upward reflection of the image at position $Y^{(0)}_3 = -y_3$, located a distance $\lambda = -(H+y_3)$ from the top surface, creates a new image at position $y^{(-1)}_3 = 2H+y_3$.
Its image tensor is given by applying the T operator linearly to all Stokeslets in the image,
\begin{align}
\mathcal{G}_{ij} & (\vec{x}, \vec{y}^{(-1)}) \notag \\
&= \mbox{T} ~ \mathcal{B}_{ij}(\vec{x}, \vec{Y}^{(0)}) \label{eq:secondTreflectionBimage}
\\
&= \mbox{T} ~ \left (  
(- \delta_{jk} + 2 y_3 \delta_{k3} \tilde{\partial}_j + y_3^2 \mbox{M}_{jk} \tilde{\nabla}^2) \mathcal{J}_{ik}(\vec{x}, \vec{Y}^{(0)})
\right ) \notag 
\\
&= (- \delta_{jk} + 2 y_3 \delta_{k3} \tilde{\partial}_j + y_3^2 \mbox{M}_{jk} \tilde{\nabla}^2) 
\left (  \mbox{T} ~  \mathcal{J}_{ik}(\vec{x}, \vec{Y}^{(0)}) \right ) \notag 
\\
&= (- \delta_{jk} + 2 y_3 \delta_{k3} \tilde{\partial}_j + y_3^2 \mbox{M}_{jk} \tilde{\nabla}^2) 
\nonumber \\
&
\quad\,\Big(
(- \delta_{kl} - 2 (H+y_3) \delta_{l3} \mbox{M}_{ku} \tilde{\partial}_u \notag
\\
&\quad\,+ (H+y_3)^2 \mbox{M}_{kl} \tilde{\nabla}^2)
\mathcal{J}_{il}(\vec{x}, \vec{y}^{(-1)})
\Big) \, .\notag
\end{align}
The replacement rule is listed as the 3rd entry in \tbl{tab:images2} and the final expression as the 3rd entry in \tbl{tab:imagesExplicit}. 
This expression may be verified by adding (\ref{eq:blakeTensor1}) to (\ref{eq:secondTreflectionBimage}) and ascertain that the no-slip condition holds on $x_3 = H$ for all $i,j$.

Similarly, the higher order image tensors are found by recursively applying the reflection operations,
\begin{eqnarray}
\label{eq:imageReplacementRule1}
\mathcal{G}_{ij}(\vec{x}, \vec{Y}^{(m)}) &=& \mbox{B} \; \mathcal{G}_{ij}(\vec{x}, \vec{y}^{(-m)}) \, , \\
\label{eq:imageReplacementRule2}
\mathcal{G}_{ij}(\vec{x}, \vec{Y}^{(-m)}) &=& \mbox{T} \; \mathcal{G}_{ij}(\vec{x}, \vec{y}^{(m-1)}) \, , \\
\label{eq:imageReplacementRule3}
\mathcal{G}_{ij}(\vec{x}, \vec{y}^{(-m)}) &=& \mbox{T} \; \mathcal{G}_{ij}(\vec{x}, \vec{Y}^{(m-1)}) \, , \\
\label{eq:imageReplacementRule4}
\mathcal{G}_{ij}(\vec{x}, \vec{y}^{(m)}) &=& \mbox{B} \; \mathcal{G}_{ij}(\vec{x}, \vec{Y}^{(-m)}) \, , 
\end{eqnarray}
where $m\ge1$. 
These replacement rules are written out for the first few images in \tbl{tab:images2}, and \tbl{tab:imagesExplicit} gives the resulting expressions of the image tensors explicitly.

Finally, adding all images together we obtain the Green's function for a Stokeslet between two parallel no-slip surfaces,
\begin{eqnarray}
v_i(\vec{x}, \vec{y}, \vec{f}) &=& \mathcal{F}_{ij} f_j \, , \, \, 
\\
\label{eq:fundamentalTensorInFilm}
\mathcal{F}_{ij}(\vec{x}, \vec{y}) &=& \sum_{m=-\infty}^{\infty} \left[ \mathcal{G}_{ij}(\vec{x}, \vec{y}^{(m)}) + \mathcal{G}_{ij}(\vec{x}, \vec{Y}^{(m)}) \right] \, . \quad\quad
\end{eqnarray}
Note that the no-slip condition can be satisfied exactly on the bottom surface by adding up the reflections to $n = 1,5,9,13, \dots$ from \tbl{tab:imagesExplicit}, and satisfied exactly on the top surface by adding up the reflections to $n = 3,7,11,15, \dots$. 
However, if symmetric flow fields are required about the channel centerline, an even number of images $n = 2,4,6,8, \dots$ must be employed.

\begin{table*}
\begin{footnotesize}
\begin{tabular}{|l||l|}
\hline
(n) & Image system tensor $\mathcal{G}_{ij}(\vec{x}, \vec{y}^{(m)} \mbox{ or } \vec{Y}^{(m)}) =$\\ \hline
(0) 	 
& $\mathcal{J}_{ij}(\vec{x}, \vec{y}^{(0)})$  
\\ 
(1) 	
&$(- \delta_{jk} + 2 y_3 \delta_{k3} \tilde{\partial}_j + y_3^2 \mbox{M}_{jk} \tilde{\nabla}^2) 
\mathcal{J}_{ik}(\vec{x}, \vec{Y}^{(0)}) $
\\ 
(2) 	
& $(- \delta_{jk} - 2 (H-y_3) \delta_{k3} \tilde{\partial}_j + (H-y_3)^2 \mbox{M}_{jk} \tilde{\nabla}^2) 
\mathcal{J}_{ik}(\vec{x}, \vec{Y}^{(-1)})$     
\\  
(3) 	
& $(- \delta_{jk} + 2 y_3 \delta_{k3} \tilde{\partial}_j + y_3^2 \mbox{M}_{jk} \tilde{\nabla}^2) $
$(- \delta_{kl} - 2 (H+y_3) \delta_{l3} \mbox{M}_{ku} \tilde{\partial}_u+ (H+y_3)^2 \mbox{M}_{kl} \tilde{\nabla}^2)
\mathcal{J}_{il}(\vec{x}, \vec{y}^{(-1)}) $
\\ 
(4) 	
& $(- \delta_{jk} - 2 (H-y_3) \delta_{k3} \tilde{\partial}_j + (H-y_3)^2 \mbox{M}_{jk} \tilde{\nabla}^2)$
$ (- \delta_{kl} + 2 (2H-y_3) \delta_{l3} \mbox{M}_{ku} \tilde{\partial}_u + (2H-y_3)^2 \mbox{M}_{kl} \tilde{\nabla}^2) 
\mathcal{J}_{il}(\vec{x}, \vec{y}^{(1)})$
\\ 
(5) 	
& $(- \delta_{jk} + 2 y_3 \delta_{k3} \tilde{\partial}_j + y_3^2 \mbox{M}_{jk} \tilde{\nabla}^2)$
$(- \delta_{kl} - 2 (H+y_3) \delta_{l3} \mbox{M}_{ku} \tilde{\partial}_u+ (H+y_3)^2 \mbox{M}_{kl} \tilde{\nabla}^2)$
\\ &
$(- \delta_{lo} + 2 (2H+y_3) \delta_{o3} \tilde{\partial}_l + (2H+y_3)^2 \mbox{M}_{lo} \tilde{\nabla}^2) 
\mathcal{J}_{io}(\vec{x}, \vec{Y}^{(1)})$
\\ 
(6) 	
& 
$(- \delta_{jk} - 2 (H-y_3) \delta_{k3} \tilde{\partial}_j + (H-y_3)^2 \mbox{M}_{jk} \tilde{\nabla}^2)$
$(- \delta_{kl} + 2 (2H-y_3) \delta_{l3} \mbox{M}_{ku} \tilde{\partial}_u + (2H-y_3)^2 \mbox{M}_{kl} \tilde{\nabla}^2)$
\\ &
$ (- \delta_{lo} - 2 (3H-y_3) \delta_{o3} \tilde{\partial}_l + (3H-y_3)^2 \mbox{M}_{lo} \tilde{\nabla}^2)
\mathcal{J}_{io}(\vec{x}, \vec{Y}^{(-2)})$
\\ 
(7) 	
& $(- \delta_{jk} + 2 y_3 \delta_{k3} \tilde{\partial}_j + y_3^2 \mbox{M}_{jk} \tilde{\nabla}^2) $
$(- \delta_{kl} - 2 (H+y_3) \delta_{l3} \mbox{M}_{ku} \tilde{\partial}_u+ (H+y_3)^2 \mbox{M}_{kl} \tilde{\nabla}^2)$
\\ &
$ (- \delta_{lo} + 2 (2H+y_3) \delta_{o3} \tilde{\partial}_l + (2H+y_3)^2 \mbox{M}_{lo} \tilde{\nabla}^2) $
$(- \delta_{op} - 2 (3H+y_3) \delta_{p3} \mbox{M}_{ov} \tilde{\partial}_v + (3H+y_3)^2 \mbox{M}_{op} \tilde{\nabla}^2)
 \mathcal{J}_{ip}(\vec{x}, \vec{y}^{(-2)})$
\\ 
(8) 	
& 
$(- \delta_{jk} - 2 (H-y_3) \delta_{k3} \tilde{\partial}_j + (H-y_3)^2 \mbox{M}_{jk} \tilde{\nabla}^2)$
$(- \delta_{kl} + 2 (2H-y_3) \delta_{l3} \mbox{M}_{ku} \tilde{\partial}_u + (2H-y_3)^2 \mbox{M}_{kl} \tilde{\nabla}^2)$
\\ &
$ (- \delta_{lo} - 2 (3H-y_3) \delta_{o3} \tilde{\partial}_l + (3H-y_3)^2 \mbox{M}_{lo} \tilde{\nabla}^2)$
$ (- \delta_{op} + 2 (4H-y_3) \delta_{p3} \mbox{M}_{ov} \tilde{\partial}_v + (4H-y_3)^2 \mbox{M}_{op} \tilde{\nabla}^2)
\mathcal{J}_{ip}(\vec{x}, \vec{y}^{(2)})$
\\ 
~ \vdots & ~ \vdots      \\ \hline
\end{tabular}
\end{footnotesize}
\caption{
Explicit expressions of the image system tensors $\mathcal{G}_{ij}$ of the first few image systems of a Stokeslet between two parallel no-slip walls.
The indices $i,j,k,l,o,p,u,v \in \{1,2,3\}$, and repeated indices are summed over. 
Added together, these tensors yield the Green's function of flow between two parallel no-slip walls.
}
\label{tab:imagesExplicit}
\end{table*}


\section{Flow far-field of a three-sphere swimmer in bulk}\label{appendix:3sphereSwimmer}

In this appendix, we show how the flow-far field of a three-sphere swimmer can well be described by a combination of dipolar and quadrupolar flows.
In the particular situation of the internal forces symmetrically distributed along the swimming axis, the dipolar contribution vanishes since the swimmer becomes invariant under time-reversal and parity transformation~\cite{pooley07}.
This type of swimmer is referred to as a self-T-dual swimmer whose leading term in the flow-far field is a quadrupole.

Firstly, we assume that the spheres have the same radius~$a$ but different oscillation amplitudes.
The flow-far field is given by~\cite{alexander09}
\begin{equation}
	\begin{split}
		\vv =& \alpha V \left(\frac{a}{s}\right)^2 
			\left( 3\left( \vect{\hat{t}} \cdot \vect{\hat{s}} \right)^2-1 \right) \vect{\hat{s}} \\
			&+ \sigma V \left( \frac{a}{s} \right)^3
			\bigg[ 3\left( 5 \left( \vect{\hat{t}}\cdot \vect{\hat{s}} \right)^3-3\left( \vect{\hat{t}}\cdot \vect{\hat{s}} \right) \right) \vect{\hat{s}} \\
			&- \left( 3\left( \vect{\hat{t}}\cdot \vect{\hat{s}} \right)^2-1 \right) \vect{\hat{t}}	\bigg] + \bigO \left( \frac{1}{s^4} \right) \, , 
	\end{split}
	\label{far-field-three-sphere-swimmer}
\end{equation}
where the unit vector $\vect{\hat{s}}:=\vect{s}/s$ and the leading order swimming velocity averaged over one period is $V=-\frac{7}{24} \, aK$.
In addition, the dipolar and quadrupolar coefficients are given by
\begin{equation}
	\alpha = \frac{27}{56} \frac{u_{20}^2-u_{10}^2}{a} \, , \qquad
	\sigma = \frac{15}{56} \frac{1}{a^2} \, .
\end{equation}

While the quadrupolar coefficient takes only positive values, the dipolar coefficient can be of different signs depending on the difference in the amplitude of the oscillations.
If $|u_{20}| > |u_{10}|$, then the dipole coefficient is positive, $\alpha>0$, and thus the swimmer is a pusher that pushes out the fluid along its swimming axis.
In contrast to that, if $|u_{20}| < |u_{10}|$, the swimmer is a puller as it pulls the fluid inward along its swimming path.

It is worth noting that the aforementioned assumption $2a+|u_{10}|+|u_{20}| \ll L$ yields that $\alpha$ is necessarily much smaller than~$\sigma$.
Accordingly, the ratio between the dipolar and quadrupolar coefficients in absolute value $|\alpha/\sigma| = \frac{9}{5} \, a | u_{20}^2-u_{10}^2 |$ can be even three orders of magnitude smaller than 1.
For instance, by taking $u_{10}=a=0.1$ and $u_{20}=2u_{10}$, the ratio $|\alpha/\sigma| = 5.4 \times 10^{-3}$.
Even though the dipolar term persists for $u_{10} \ne u_{20}$, the flow field is primly dominated by the quadrupolar contribution, at intermediate distances from the swimmer.

We next assume that $u_{10}=u_{20}$ and consider the case in which the spheres have different sizes, as is considered in the present work.
The swimming velocity averaged over one full cycle reads~\cite{Golestanian2008a}
\begin{equation}
	V = -\frac{21K}{8} \frac{a_1 a_2 a_3}{\left(a_1+a_2+a_3\right)^2} \, . 
\end{equation}

In addition, the dipolar and quadrupolar coefficients of the corresponding flow field are given by
\begin{equation}
	\alpha = \frac{3}{4} \frac{a_2-a_3}{a^2} \, , \qquad
	\sigma = \frac{3}{56} \frac{4\left(a_2+a_3\right)-3a}{a^3} \, ,
	\label{alpha-sigma-unequal-radii}
\end{equation}
where $a$ is taken as the radius of the central sphere $a_1$.
Remarkably, the swimmer is a pusher (puller) if $a_2>a_3$ $(a_2<a_3)$, independently of the central sphere~$a$.
In addition, if $a < \frac{4}{3} \left(a_2+a_3\right)$, then the quadrupolar coefficient is positive, $\sigma>0$, a situation which characterizes swimmers with small bodies and elongated flagella. 
The flow-far field of the swimmer can be dipolar- or quadrupolar-dominated at intermediate distances from the swimmer, depending on the sizes of the spheres.

We finally assess the effect of the mean arm lengths on the far-field hydrodynamics.
By posing $L_2=\beta L_1 = L$ and scaling the lengths by $L$, the averaged swimming velocity is given by~\cite{Golestanian2008a}
\begin{equation}
	V = -\frac{aK}{6} \left( 1 + \frac{1}{\beta^2}-\frac{1}{\left(1+\beta\right)^2} \right) \, , 
\end{equation}
and the dipolar and quadrupolar moments follows as
\begin{equation}
	\alpha = \frac{3}{8a} \frac{N_\alpha}{D} \left(1-\beta\right) \, , \qquad 
	\sigma = \frac{3}{16a^2} \frac{N_\sigma}{D} \, , 
\end{equation}
where we have defined for convenience the quantities
\begin{align}
	N_\alpha &= \beta \left( 2+7\beta+11\beta^2+7\beta^3+2\beta^4 \right) \, , \notag \\
	N_\sigma &= \beta \big(2+4\beta+\beta^2-4\beta^3+\beta^4 
	+ 4\beta^5+2\beta^6\big) \, , \notag \\
	D &= 1+2\beta+\beta^2+2\beta^3+\beta^4 \, . \notag
\end{align}

The swimmer is a pusher (puller) if $\beta<1$ $(\beta>1)$.
Moreover, $\sigma>0$ for all positive values of the parameter $\beta$.



\section{Mathematical formulas}\label{appendix:formulas}

In this appendix, we provide explicit analytical expressions of the functions and coefficients stated in the main text.

\subsection{Expressions of $A(z)$, $B(z)$ and $C(z)$ for a neutral swimmer (equal sphere radii)}\label{subappendix:ABC}

\begin{center}

		\def\arraystretch{1.8}
			\begin{table*}
					\begin{tabular}{|c|c|}
							\hline
							$A_0$ & $\tfrac{7}{24}- \tfrac{7}{24} w_2w_1 -\tfrac{w_2}{24} +\tfrac{w_1}{3}$ \\ 
							\hline
							$A_2$ & $-{\tfrac {35}{4}} \,w_2w_1-\tfrac {13}{12} \, w_2+\tfrac{14}{3} \,w_1$ \\
							\hline
							$A_4$ & $-\tfrac {931}{8} \,w_2w_1-{\tfrac {1159}{96}} \, w_2+{\tfrac {91}{3}} \,w_1$ \\
							\hline
							$A_6$ & $-\tfrac {5425}{6} \, w_2w_1-{\tfrac {2407}{32}} \, w_2+172 \, w_1$ \\
							\hline
							$A_8$ & $-{\tfrac {36435}{8}} \, w_2w_1-{\tfrac {9203}{32}} \, w_2+977 \, w_1$ \\
							\hline
							$A_{10}$ & $-{\tfrac {62475}{4}} \, w_2w_1-{\tfrac {22597}{32}} \, w_2+4042 \, w_1$ \\
							\hline
							$A_{12}$ & $-{\tfrac {298445}{8}} \, w_2w_1-{\tfrac {4721}{4}} \, w_2+10567 \, w_1$ \\
							\hline
							$A_{14}$ & $-62475 \, w_2w_1-{\tfrac {3107}{2}} \, w_2+17312 \, w_1$ \\
							\hline
							$A_{16}$ & $-72870 \, w_2w_1-1904 \, w_2+17812 \, w_1 $ \\
							\hline
							$A_{18}$ & $-\tfrac {173600}{3} \, w_2w_1-\tfrac {5768}{3} \, w_2+\tfrac {33616}{3} \, w_1$ \\
							\hline
							$A_{20}$ & $-29792 \, w_2w_1-{\tfrac {3520}{3}} \, w_2+{\tfrac {11840}{3}} \, w_1$ \\
							\hline
							$A_{22}$ & $-8960 \, w_2w_1-{\tfrac {896}{3}} \, w_2+{\tfrac {1792}{3}} \, w_1 $ \\
							\hline
							$A_{24}$ & $-{\tfrac {3584}{3}} \, w_2 w_1$ \\
							\hline
							\hline
						\end{tabular}
	\quad
		\def\arraystretch{1.68}
					\begin{tabular}{|c|c|}
			\hline
			$B_{-1}$ & ${\tfrac {3}{32}} \, w_2w_1$ \\
			\hline
			$B_0$ & $\tfrac{7}{24} -{\tfrac {7}{24}} \, w_2w_1-\tfrac{w_2}{24} +\tfrac{w_1}{3}$ \\
			\hline
			$B_1$ & $\tfrac {45}{32} \, w_2w_1$ \\
			\hline
			$B_2$ & $-{\tfrac {35}{8}} \, w_2w_1-{\tfrac{89}{96}} \, w_2+\tfrac{8}{3} \, w_1$ \\
			\hline
			$B_3$ & ${\tfrac {261}{32}} \, w_2w_1 $ \\
			\hline
			$B_4$ & $-{\tfrac {203}{8}} \, w_2w_1-{\tfrac {175}{24}} \, w_2+{\tfrac {82}{3}} \, w_1$ \\
			\hline
			$B_5$ & ${\tfrac {735}{32}} \, w_2w_1$  \\
			\hline
			$B_6$ & $-{\tfrac {1715}{24}} \, w_2w_1-{\tfrac {2425}{96}} \, w_2+{\tfrac {208}{3}} \, w_1$  \\
			\hline
			$B_7$ & ${\tfrac {261}{8}} \, w_2w_1$  \\
			\hline
			$B_8$ & $-{\tfrac {203}{2}} \, w_2w_1-{\tfrac {875}{24}} \, w_2+{\tfrac {197}{3}} \, w_1$  \\
			\hline
			$B_9$ & ${\tfrac {45}{2}} \, w_2w_1$  \\
			\hline
			$B_{10}$ & $-70\, w_2w_1-{\tfrac {91}{6}} \, w_2+{\tfrac {64}{3}} \, w_1$  \\
			\hline
			$B_{11}$ & $6 \, w_2w_1  $\\
			\hline
			$B_{12}$ & $-{\tfrac {56}{3}} \, w_2w_1-6w_2$ \\
			\hline
			\hline
					\end{tabular}
	\quad
		\def\arraystretch{2.155}
					\begin{tabular}{|c|c|}
			\hline
			$C_3$ & $\tfrac{3}{16} \, w_2-6 \, w_1$  \\
			\hline
			$C_5$ & ${\tfrac {321}{64}} \, w_2-102 \, w_1$  \\
			\hline
			$C_7$ & ${\tfrac {3699}{64}} \, w_2-702 \, w_1$  \\
			\hline
			$C_9$ & ${\tfrac {23931}{64}}\, w_2-2502 \, w_1$  \\
			\hline
			$C_{11}$ & ${\tfrac {94869}{64}} \, w_2-4842 \, w_1$  \\
			\hline
			$C_{13}$ & ${\tfrac {29457}{8}} \, w_2-4482 \, w_1 $ \\
			\hline
			$C_{15}$ & ${\tfrac {22419}{4}} \, w_2+198 \, w_1  $\\
			\hline
			$C_{17}$ & $4824 \, w_2+4878 \, w_1  $\\
			\hline
			$C_{19}$ & $1836 \, w_2+4968 \, w_1 $ \\
			\hline
			$C_{21}$ & $-96 \, w_2+2208 \, w_1  $ \\
			\hline
			$C_{23}$ & $-192 \, w_2+384 \, w_1$ \\
			\hline
			\hline
					\end{tabular}
			\caption{The coefficients $A_n$, $B_n$ and $C_n$ of the series functions defined in Eq.~\eqref{seriesFctsABC}.
			Here $w_1 := \sqrt{1+z^2}$ and $w_2 := \sqrt{1+4z^4}$.
			}
			\label{table:seriesCoeff}
			\end{table*}
		
\end{center}

Here we provide explicit analytical expressions of the functions $A(z)$, $B(z)$, and $C(z)$ defined in Eq.~\eqref{swimmingTrajABC_neutral} of the main text, to leading order in $a$ and as a power series in $z$.
Defining $w_1 := \sqrt{1+z^2}$ and $w_2 := \sqrt{1+4z^4}$, we have
\begin{subequations}
	\begin{align}
		A(z) &= \frac{a}{(w_1w_2)^{13}} \sum_{n=0}^{12} A_{2n} z^{2n} \, , \\
		B(z) &= \frac{a}{(w_1w_2)^{7}} \sum_{n=-1}^{12} B_{n} z^{n} \, , \\
		C(z) &= \frac{a}{(w_1w_2)^{13}} \sum_{n=1}^{11} C_{2n+1} z^{2n+1} \, .
	\end{align}\label{seriesFctsABC}
\end{subequations}

The series coefficients $A_n$, $B_n$, and $C_n$ are given in Tab.~\ref{table:seriesCoeff}.


\allowdisplaybreaks
\subsection{Analytical expressions for a general three-sphere swimmer in the far-field limit}\label{subappendix:generalSwimmerFarField}


The explicit analytical expression of the coefficients $V_{10}$ and $V_{20}$ defined in Eq.~\eqref{bulkSwimmingVelocityGeneral} are
\begin{equation}
		V_{10} = -\frac{21PK}{8M^2} \, , \quad
		V_{20} = \frac{9 PK\left( 18-27S-6P+11Q \right)}{32M^3}  \, . \notag 
\end{equation}

The coefficients $A_{ij}$, $B_{ij}$, $C_{ij},$ and $D_{ij}$ defined in Eqs.~\eqref{expr_ABC_general} and \eqref{expr_D_general} are given by
\begin{align}
	A_{23} &= -\frac{63P \left( 9+25S+88P-12Q \right)}{1024 M^3}  \, , \notag \\
	D_{14} &= \frac{135P}{64 MN}  \, , \notag \\
	D_{22} &= -\frac{189P}{256 M^2} \, , \notag \\
	D_{24} &= \frac{135 P\left( 85Q-18S+542PS+140P+640P^2 \right)}{2048 M^2N^2}  \, , \notag \\
	B_{13} &= \frac{63P}{64M} \, , \notag \\
	B_{23} &= -\frac{63P \left(9-6PS-3Q+13S+16P\right) }{256 M^3} \, , \notag \\
	C_{14} &= \frac{405P}{256N} \, , \notag \\
	C_{24} &= \frac{405P \left(10P+13S\right)}{2048MN}  \, , \notag
\end{align}
where $S=r_{3}+r_{2}$, $P=r_{3}r_{2}$, $Q=r_{3}^2+r_{2}^2$, $M=1+S$ and $N=S+4P$.
We recall that $r_{2}=a_2/a_1$ and $r_{3}=a_3/a_1$.




%

\end{document}